\documentclass[useAMS,usenatbib]{mn2e}

\usepackage{soul}
\usepackage{tabularx}
\usepackage{xspace}
\usepackage{aas_macros}
\usepackage{array}
\usepackage{url}
\usepackage{amsmath}
\usepackage{amssymb}
\usepackage[noabbrev, capitalise]{cleveref}
\usepackage{color,graphicx}

\newcommand{\mhalo}{M_{\rm h}}

\def\simlt{\mathrel{\rlap{\lower 3pt\hbox{$\sim$}}\raise 2.0pt\hbox{$<$}}}
\def\simgt{\mathrel{\rlap{\lower 3pt\hbox{$\sim$}} \raise
2.0pt\hbox{$>$}}}
\title[The T-RECS simulation]{The Tiered Radio Extragalactic Continuum Simulation (T-RECS)}
\author[Bonaldi et al.]{Anna Bonaldi\textsuperscript{\thanks{E-mail: a.bonaldi@skatelescope.org
\newline
code: https://github.com/abonaldi/TRECS.git
\newline
\hspace{1cm} catalogues: at CDS via anonymous ftp to cdsarc.u-strasbg.fr (130.79.128.5) or via http://cdsarc.u-strasbg.fr/viz-bin/qcat?VII/282}$^1$}, Matteo Bonato$^{2,3}$, Vincenzo Galluzzi$^4$,
Ian Harrison$^5$, 
 \newauthor Marcella Massardi$^2$, Scott Kay$^5$, Gianfranco De Zotti$^3$, Michael L. Brown$^5$ \\%}
$^1$SKA Organization, Jodrell Bank, Lower Whitington, Macclesfield, SK11 9DL, UK\\
$^2$INAF$-$Istituto di Radioastronomia, and Italian ALMA Regional Centre, Via Gobetti 101, I-40129, Bologna, Italy\\
$^3$INAF$-$Osservatorio Astronomico di Padova, Vicolo Osservatorio 5, I-35122, Padova, Italy\\
$^4$INAF$-$Osservatorio Astronomico di Trieste, Via Tiepolo 11, I-34143, Trieste, Italy\\
$^5$Jodrell Bank Centre for Astrophysics, School of Physics \& Astronomy, The University of Manchester, Manchester M13 9PL, UK
}

\begin{document}
%\date{Accepted 1988 December 15. Received 1988 December 14; in original form 1988 October 11}

\pagerange{\pageref{firstpage}--\pageref{lastpage}} \pubyear{2017}

\maketitle

\label{firstpage}

\begin{abstract}
We present the Tiered Radio Extragalactic Continuum Simulation (T-RECS): a new simulation of the radio sky in continuum, over the 150\,MHz--20\,GHz range. T-RECS models two main populations of radio galaxies: Active Galactic Nuclei (AGNs) and Star-Forming Galaxies (SFGs), and corresponding sub-populations. %following \cite{Bonato2017}. 
Our model also includes polarized emission over the full frequency range, which has been characterised statistically for each population using the available information.  %\citep[e.g.,][]{Hales2014,galluzzi2018,sun_reich2012}.
We model the clustering properties in terms of probability distributions of hosting halo masses, and use lightcones extracted from a high-resolution cosmological simulation to determine the positions of haloes. This limits the sky area for the simulations including clustering to a 25\,deg$^2$ field of view. We compare luminosity functions, number counts in total intensity and polarization, and clustering properties of our outputs to up-to-date compilations of data and find a very good agreement. We deliver a set of simulated catalogues, as well as the code to produce them, which can be used for simulating observations and predicting results from deep radio surveys with existing and forthcoming radio facilities, such as the Square Kilometre Array (SKA).
\end{abstract}

\section{Introduction}
The last decade has seen a steady progress in our understanding of the radio sky. Deeper and wider radio surveys \citep[e.g.,][]{2010ApJS..188..178M, 2011AJ....142....3H, 2012ApJ...758...23C,Vernstrom2016,2017MNRAS.471..210G,Smolcic2017a} have provided key information for the modelling of the sub-mJy populations, where the emission due to star formation dominates over that of radio-loud (RL) active nuclei. Meanwhile, a relentless modelling effort has improved the interpretation of these data, and built stronger links with those at other wave-bands \citep{massardi2010,Cai2013,2014MNRAS.442..682M,2015ApJ...807..141P,2015MNRAS.447.3442L,2015MNRAS.447..168R,mancuso2015,magnelli2015,Bonato2017}. The polarization of samples of radio sources have also been targeted by recent observations, thus providing a starting point for their characterization \citep{test,Grant2010,Hales2014,2014MNRAS.444..700B,2016ApJ...829....5L,Massardi2013,Galluzzi2017,galluzzi2018}.  

The advent of the Square Kilometre Array (SKA) over the next decade will achieve a significant step beyond the current state-of-art radio observatories. The current models will be undoubtely challenged once SKA observations will be available. Until that happens, however, they are our best means to predict what SKA observations may look like. Such predictive capabilities are particularly important (i) to best design surveys that meet the various scientific objectives; (ii) to understand the computational and the data analysis challenges posed by the new observations; and (iii) to test/demonstrate the validity of ideas and approaches being developed for the SKA.

The T-RECS simulation has been developed as a way to enable these objectives. This effort is close in spirit to that of the widely-used S3-SEX simulation by \cite{wilman2008}, and it is motivated by the need for an update after ten very prolific years in terms of radio observations and modelling. T-RECS also includes polarization information for all radio sources, which is relevant for several planned SKA surveys but wasn't modelled in S3-SEX due to the lack of data at the time. It also features a realistic treatment of the clustering properties of radio sources, by associating them to Dark Matter (DM) haloes of a cosmological simulation. T-RECS predictions should hold from 150\,MHz to 20\,GHz. 

We model the radio sky in terms of two main populations: AGNs and SFGs. Recent studies \citep[e.g.,][]{kellermann2016,Padovani2015,mancuso2017,white2017} suggest the existence of a third population, of radio-quiet (RQ) AGNs. Those are star-forming galaxies that obey the relation between star formation rate and radio emission, but also host an active nucleus that contributes to the radio emission. The processes responsible for the emission of RQ AGNs, as well as their dichotomy with the RL population, are still hotly debated. At the core of the problem there is the understanding of the interaction between black-hole accretion and star-formation in galaxies, which in turn constrains the relative emission levels. 

In T-RECS there is no explicit modelling of RQ active nuclei. However, the AGN population in our model essentially maps RL AGNs, where the accretion is by far the dominant source of radio emission. Therefore, RQ AGNs would contribute part of the flux of those sources that, in T-RECS, are modelled as SFGs. As a future upgrade of our model, we plan to characterize this component explicitly, following for example \cite{mancuso2017}. 

The structure of the paper is as follows: in Sect.~\ref{sec:dm_sim} we describe the cosmological simulation used as a base for the clustering model; in Secs.~\ref{sec:AGN_model} and \ref{sec:SFG_model} we describe the AGN and the SFG models, respectively. In Sect. \ref{sec:validation} we compare our outputs with the most recent compilation of data; in Sec \ref{sec:cats} we describe our output catalogues; finally, we present our conclusions and discuss the prospects for future updates in Sect.~\ref{sec:conclu}. 

\section{Base cosmological dark matter simulation}\label{sec:dm_sim}
In order to get realistic clustering
properties for the AGNs and SFGs in our model, their positions on the sky are linked to those
of Dark Matter (DM) haloes of a cosmological simulation. We relied on the 
P-Millennium simulation \citep{baugh2017}, a DM-only simulation with a {\it
Planck} best-fitting cosmology: $H_0=67.77$\,km \,s$^{-1}$\,Mpc$^{-1}$,
$\Omega_\Lambda=0.693$, $\Omega_{\rm M}$=0.307, $\sigma_8=0.8288$
\citep{planck2014}. Initial conditions were generated at a redshift of $z=127$,
and 272 snapshots were created down to $z=0$. To generate merger trees, a
friend-of-friend algorithm was run to identify haloes and {\tt subfind}
\citep{springel2001} to identify subhaloes. Finally, subhaloes were tracked
between output times and consistently assigned memberships as described in
\cite{jiang2014}.

This simulation has been chosen as a basis for T-RECS because of
the very high particle resolution (each particle has mass $1.061 \times
10^8\,h^{-1}\,M_{\odot}$) which allows associating galaxies to individual
haloes. As a drawback, this simulation has a relatively small box size of $800$\,(Mpc/h)$^{3}$, 
which allows us to have a maximum field of view of
$5\times5$\,deg$^2$ out to the maximum redshift we considered for clustering
($z=8$). 

We generated the $5\times5$\,deg$^2$ lightcone for $z=0$--8  from the merger
tree outputs. This redshift range is sampled by 201 snapshots, identified by
their redshift $z_{\rm s}$. For each $z_{\rm s}$, given the base cosmology, we
compute the dimension perpendicular to the line of sight $\Delta_{\perp}(z_{\rm
s})$ as the comoving size corresponding to 5\,deg, and the dimension parallel
to the line of sight $\Delta_{\parallel}(z_{\rm s})$, as the difference in
comoving distance between the redshift of the current snapshot and that of the
next one. The lightcone is finally obtained by collating slices of size
$\Delta_{\parallel}\times \Delta_{\perp} \times \Delta_{\perp}$ for all
snapshots ordered by increasing redshift.

For the dimension parallel to the line of sight we started from one edge of the
box at redshift 0 and we extracted the square of side $\Delta_{\perp}(z_{\rm
s})$ with a random central coordinate. We then stepped through slices of
$\Delta_{\parallel}(z_{\rm s})$ for each snapshot, with the same field centre, thus preserving the clustering along the redshift dimension. On those
occasions where we reached the end face of the box (this happened 13 times
within the redshift range considered), we started the next snapshot from the
front face, but generated a new set of coordinates for the centre of the field
of view, thus avoiding structures repeating with redshift.

The position of the DM haloes in the cosmological simulation is
identified by three cartesian spatial coordinates (in units of Mpc/h), which we
converted to redshift and angular coordinates on the sky. For the former, we
used the redshift of the slice in the lightcone, $z_{\rm s}$, plus a correction
to take into account the position of the halo with respect to the centre of the
slice. To generate the coordinates on the sky, we first converted linearly
each of the other two halo coordinates to angles ranging between -2.5 and +2.5
degrees for each redshift slice. These angular coordinates, which we called
\emph{x\_coord} and \emph{y\_coord}, are therefore cartesian coordinates on a
plane, and they do not correspond to any set of astronomical coordinates on the
full sky.  However, the $5^\circ \times 5^\circ$ sky area considered is small
enough that the portion of sphere can be approximated with a plane. In this
case, once a central coordinate for the field is specified, the x\_coord and y\_coord 
coordinates are projected to the spherical ones, \emph{longitude} and \emph{latitude}. 
The catalogues are delivered with (0,0) central spherical coordinates, but code is provided 
to project x\_coord and y\_coord easily to another direction in the sky. 

The haloes in the lightcone have been associated to AGNs and SFGs with
different methods, described in sections \ref{sec:cluster_agn} and
\ref{sec:cluster_sfg}, respectively. The halo mass distributions for the two
populations derived from those analyses turn out to be quite different, with RL
AGNs being typically associated with higher masses than SFGs. This results in different 
clustering properties, which are compared to the data in Sect. \ref{sec:valid_clust}.

\section{Active Galactic Nuclei model description}\label{sec:AGN_model}
\subsection{Base evolutionary model}
To describe the cosmological evolution of the luminosity function (LF) of
RL AGNs we adopted an updated version of the \citet{massardi2010}
model, slightly revised by \citet{Bonato2017}. 

The best-fit values of the parameters were re-computed adding to the fitted data sets the 4.8\,GHz number counts for the flat-spectrum population by \citet{Tucci2011}. This addition has resulted in a significant improvement of the evolutionary model for flat-spectrum sources, while affecting only marginally that for the steep-spectrum population.

The model comprises three
source populations with different evolutionary properties: steep-spectrum
sources (SS-AGNs), flat-spectrum radio quasars (FSRQs) and BL Lacs. For sources
of each population, \cite{Bonato2017} adopts a simple power-law spectrum: $S\propto
\nu^{\alpha}$, with $\alpha_{\rm FSRQ}=\alpha_{\rm BLLac}=-0.1$, and
$\alpha_{\rm steep} = -0.8$.

The epoch-dependent comoving LFs (in units of ${\rm Mpc^{-3}\, (d\log
L)^{-1}}$) are modeled as double power-laws:
\begin{equation}\label{eq:Phi}
\Phi(L(z), z)=\frac{n_0}{(L(0)/L_\star(0))^a+(L(0)/L_\star(0))^b}{d\log L(0)\over d\log L(z)}.
\end{equation}
The evolution with redshift of the characteristic luminosity $L_\star$ of each
population is described by the analytic formula
\begin{equation}\label{eq:evol}
L_{\star}(z)\!=\! L_{\star}(0){\rm dex}\!{\left[k_{\rm evo}z\!\left(\!2z_{\rm top}\!-\!{z^{m_{\rm ev}}z_{\rm top}^{(1-m_{\rm ev})}}\!/(1\!+\!m_{\rm ev})\right)\right]},
%
% L_\star(z) = L_{\star}(0) 10^{[k_{\rm evo} z \!(\!2z_{\rm top}\!-\!{z^{m_{\rm ev}}z_{\rm top}^{(1-m_{\rm ev})}}\!/(1\!+\!m_{\rm ev})) ]} .
%
%\log(L_{\star}(z))\!=\!\log (L_{\star}(0))\left[k_{\rm evo}z\!\left(\!2z_{\rm top}\!-\!{z^{m_{\rm ev}}z_{\rm top}^{(1-m_{\rm ev})}}\!/(1\!+\!m_{\rm ev})\right)\right]
\end{equation}
that entails a high-$z$ decline of the comoving LF. The redshift, $z_{\rm
top}$, at which $L_{\star}(z)/ L_{\star}(0)$ reaches its maximum is
luminosity-dependent
\begin{equation}\label{eq:ztop}
z_{\rm top}=z_{{\rm top},0}  + {\delta z_{\rm top}\over 1+ L_{\star}(0)/L}.
\end{equation}
This expression models the observed trend in which the high-$z$ decline of the space
density is more pronounced and starts at lower redshifts for less powerful
sources, in a way qualitatively similar to the \textit{downsizing} observed for
galaxies and optically and X-ray selected quasars \citep[see,
e.g.,][]{DeZotti2010}.

The new best fit values of the parameters of equations (\ref{eq:Phi})-(\ref{eq:ztop}) are given in Table~\ref{tab:agnparm}.
The fitted data, which include  number counts, luminosity functions and redshift distributions for flat- and steep-spectrum sources \citep[see][]{massardi2010} require a quite strong luminosity dependence of the peak redshift for the steep-spectrum population.

In the case of FSRQs the
evolution of the low luminosity portion of the LF is poorly constrained by the
data; as a result, there is only a weak evidence of a luminosity dependence of
$z_{\rm top}$ ($\delta z_{\rm top}\ll 1$). As for BL Lacs, there are not enough available data to constrain the parameters governing the luminosity dependence of the
evolution. Thus, for this population, following \citet{massardi2010},
\citet{Bonato2017} have set $m_{\rm ev}=1$ and $\delta z_{\rm top}= 0$.

We note that, in the framework of this luminosity-dependent luminosity
evolution model, the steep slope of the bright end of the LFs ($L\gg L_\star$),
particularly of FSRQs and SS-AGNs, implies strong evolution. In the case of
SS-AGNs we are in the luminosity range of FR\,II radio sources
\citep{FanaroffRiley1974}, nearly all of which have 1.4\,GHz luminosity above
$10^{25}\, \hbox{W}\,\, \hbox{Hz}^{-1}$. These sources are believed to be
typically powered by radiatively efficient accretion of cold gas from a
geometrically thin, optically thick accretion disc. This accretion produces
high-excitation emission lines; hence these objects are referred to as
high-excitation radio galaxies \citep[HERGs; e.g.,][]{McAlpine2013}.

On the contrary, the relatively flat shape of the faint end of the LFs,
particularly in the case of SS-AGNs, implies a weak evolution of sources with
$L\ll L_\star$, consistent with  the results by  \citet{McAlpine2013} and
\citet{Best2014}. These sources have luminosities in the range of FR\,I radio
sources \citep{FanaroffRiley1974}. They are currently interpreted as being
powered by radiatively inefficient accretion flows at low Eddington ratios
\citep{HeckmanBest2014}. The bulk of their energetic output is in kinetic form,
in two-sided collimated outflows (jets); they are therefore referred to as
``jet-mode'' AGNs. The strong emission lines normally found in powerful AGNs
are generally absent; they are thus also referred to as low-excitation radio
galaxies (LERGs).
\subsection{Total intensity number counts}
\begin{table}
 \caption{Best-fit values of the parameters of the evolutionary model for radio AGNs, re-computed following Bonato et al. (2017) but including an additional data set (see text). The luminosity $L_{*}$
 is in $\hbox{W}\,\hbox{Hz}^{-1}$.}
  \begin{tabular}{lccc}
    \hline
    Parameter          & FSRQ  & BLLac & SS-AGN   \\
    \hline
$a$ &  0.776   &  0.723  &  0.508 \\
$b$ &  2.669   &  1.918  &  2.545 \\
$\log n_{0}$ &  -8.319   &  -7.165  & -5.973  \\
$\log L_{*}(0)$ & 33.268    &  32.282  & 32.560  \\
$k_{\rm evo}$ &  1.234   & 0.206   & 1.349  \\
$z_{\rm top,0}$ &  2.062   &  1.262  &  1.116 \\
$\delta z_{\rm top}$ &  0.559   & $-$   &  0.705 \\
$m_{\rm ev}$ & 0.136    &  1  & 0.253  \\
    \hline
  \end{tabular}  \label{tab:agnparm}
\end{table}

The model described in the previous sub-section has been used to simulate
the number counts of AGNs at 1.4\,GHz. In practice we adopted the following
procedure. Consider a small flux density interval $\Delta S_j=S_{\rm max,
j}-S_{\rm min,j}$ and let $\Phi(L|z)$ be the luminosity function per dex (i.e.
per unit $d\log(L)$) at the redshift $z$. The contribution to the counts from
the small redshift interval $\Delta z_i$ is, approximately:
\begin{equation}\label{eq:counts_part}
\Delta N(S)_{i,j}= \Omega\, \Phi(L|z_i) \left({dV(z)\over dz}\right)_{z=z_i}\!\!\!\! \Delta \log(L_{i,j})\,\Delta z_i,
\end{equation}
where $\Omega$ is the solid angle of the simulation, $z_i$ is the center of the
redshift bin, $dV(z)/dz$ is the volume element per unit solid angle and $\Delta
\log(L_{i,j})=\log[L(S_{\rm max, j}, z_{\rm max, i})]-\log[L(S_{\rm min, j},
z_{\rm min, i})]$. Obviously the maximum and minimum values refer to the
boundaries of the corresponding bins. The total counts within $\Delta S_j$ and
$\Omega$ are then
\begin{equation}\label{eq:counts_tot}
N(S)\Delta S_j= \sum_i \Delta N(S)_{i,j}.
\end{equation}
The $N(S)\Delta S_j$ sources were then randomly distributed within the $\Delta
\log(L)$ and associated to the halos in the volume corresponding to
$\Omega\,\Delta z_i$ as specified in sub-sect.~\ref{sec:cluster_agn}. The
accuracy of this approximation was tested comparing the derived $N(S)\Delta
S_j$ with the model counts and found to be good for $\delta\log z \simeq 0.006$
and $\delta\log S=0.11$.

To make the simulations more realistic we decided to go beyond the simple
approximation of a single spectral index for all sources of each population.
The approach we have chosen also allows us to take into account systematic
variations with frequency of the spectral index distributions, clearly
demonstrated by multi-frequency observations \citep[e.g.,][]{Bonavera2011,
Bonaldi2013, Massardi2011, Massardi2016}. The effective spectral index between
the frequencies $\nu_1$ and $\nu_2$ of sources of a given population with flux
density $S_1$, within $dS_1$, at $\nu_1$,
\begin{equation}\label{eq:alpha_eff1}
\alpha_{\rm eff}(\nu_1,\nu_2)=\log(S_2/S_1)/\log(\nu_2/\nu_1),
\end{equation}
was computed finding the flux density $S_2$ at $\nu_2$ such as
$N_1(S_1)dS_1=N_2(S_2)dS_2$. Thus $\alpha_{\rm eff}(\nu_1,\nu_2)$ is the single
spectral index relating the counts at $\nu_1$ to those at $\nu_2$. The
differential source counts $N(S)$ at the two frequencies were obtained from
the updated \citet{massardi2010} model up to 5\,GHz
and the \citet{DeZotti2005} model at higher frequencies.

We adopted a Gaussian spectral index distribution with mean
$\alpha(\nu_1,\nu_2)$ and dispersion $\sigma$. The mean spectral index is
related to $\alpha_{\rm eff}(\nu_1,\nu_2)$ by \citep{Kellermann1964,
Condon1984, DaneseDeZotti1984}:
\begin{equation}\label{eq:alpha_eff}
\alpha_{\rm eff}(\nu_1,\nu_2)=\alpha(\nu_1,\nu_2)-\sigma^2(1-\beta)\ln(\nu_2/\nu_1),
\end{equation}
where $\beta$ is the slope of the differential number counts at $S_1$, computed
from the models. For each population, $\alpha_{\rm eff}$ is the fixed spectral
index used in the models. The dispersion was set at $\sigma=0.25$ for all
populations, consistent with the results by \citet{Ricci2006} after allowing
for the contribution of measurement errors to the observed dispersion. Then,
$\alpha(\nu_1,\nu_2)$ was obtained from eq.~(\ref{eq:alpha_eff}). As shown by
this equation, the mean spectral index varies with flux density because of the
variation of the slope, $\beta$, of the counts. If $\nu_1<\nu_2$ the effective
spectral index, $\alpha_{\rm eff}$, is larger than the mean value $\alpha$, as
a consequence of the fact that higher frequency surveys favour sources with
`harder' spectra.

The simulations cover the frequency range from 150 MHz to 20 GHz. We have taken
1.4\,GHz as our reference frequency and reached 20 GHz in two steps. First we
have computed the mean spectral indices between 1.4 and 4.8\,GHz in steps of
$\delta\log(S)=0.08$; the variations of $\beta$ over this flux density interval
are negligibly small. The  maximum variation of the mean $\alpha(1.4,4.8)$ over
the full flux density range of our simulations is $\delta\alpha\simeq 0.08$. We
have then repeated the procedure between 4.8 and 20\,GHz; in this case
$\delta\alpha\simeq 0.09$.

To each simulated source drawn from the redshift-dependent 1.4\,GHz
(rest-frame) LF of its population we have attributed a spectral index extracted at random from the Gaussian distribution with mean $\alpha(1.4, 4.8)$  and
dispersion $\sigma$ up to 4.8\,GHz, and a second spectral index extracted from
the 4.8--20\,GHz distribution up to 20\,GHz.  The 1.4--4.8\,GHz spectral index has been adopted over the whole 150\,MHz--4.8\,GHz range. As described in Sect. \ref{sec:validation}, this gives a good agreement between number counts from T-RECS and the available data across the full 150\,MHz--20\,GHz frequency range, while keeping the spectra of individual sources still relatively simple.

%FIGURES RELATED TO SECTION 5.1
\begin{figure*}
  \hspace{+0.0cm} \makebox[\textwidth][c]{
     \includegraphics[trim=4.5cm 0.0cm 5.0cm 0.cm,clip=true,width=0.9\textwidth, angle=0]{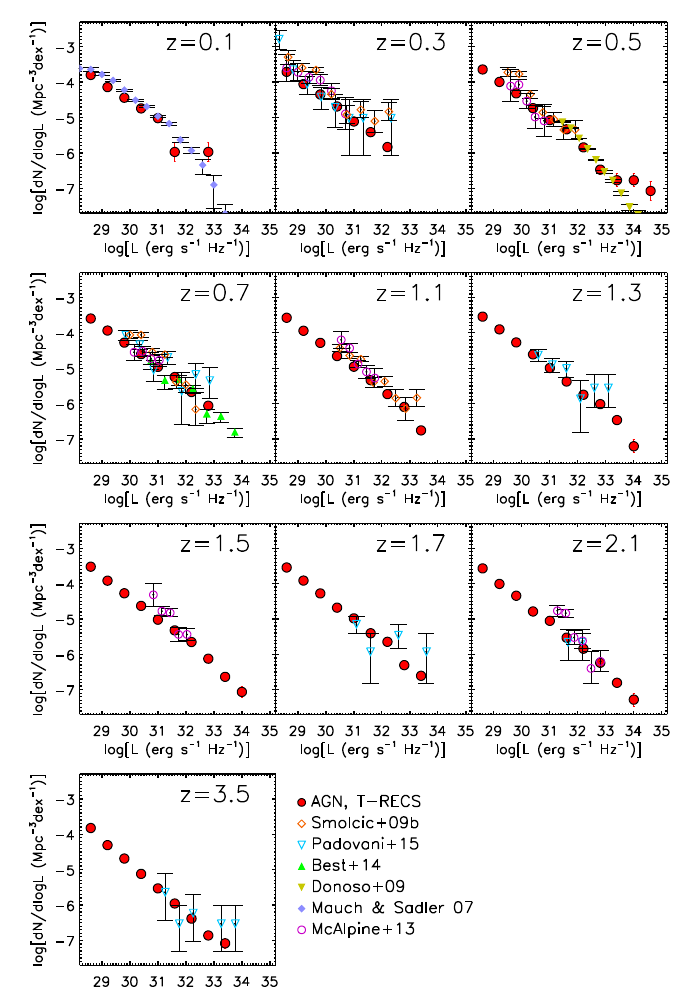}
     }
  \vspace{-0.5cm}
  \caption{Comparisons between our AGN 1.4\,GHz RLFs derived from the simulated catalogues
  (using the formalism described in Sect.~\ref{sec:AGN_model}) and observational determinations
  taken from literature (\citealt{Padovani2015}, \citealt{Best2014}, \citealt{Donoso2009},
  \citealt{MauchSadler2007}, \citealt{Smolcic2009b} and \citealt{McAlpine2013}). The simulated sky area does not always allow sampling the highest luminosities for which we have observational determinations.}
  \label{fig:1d4_LF_AGN}
\end{figure*}

\begin{figure*}
\hspace{+0.0cm} \makebox[\textwidth][c]{
  \includegraphics[trim=4.5cm 1.9cm 5.0cm 0cm,clip=true,width=0.9\textwidth, angle=0]{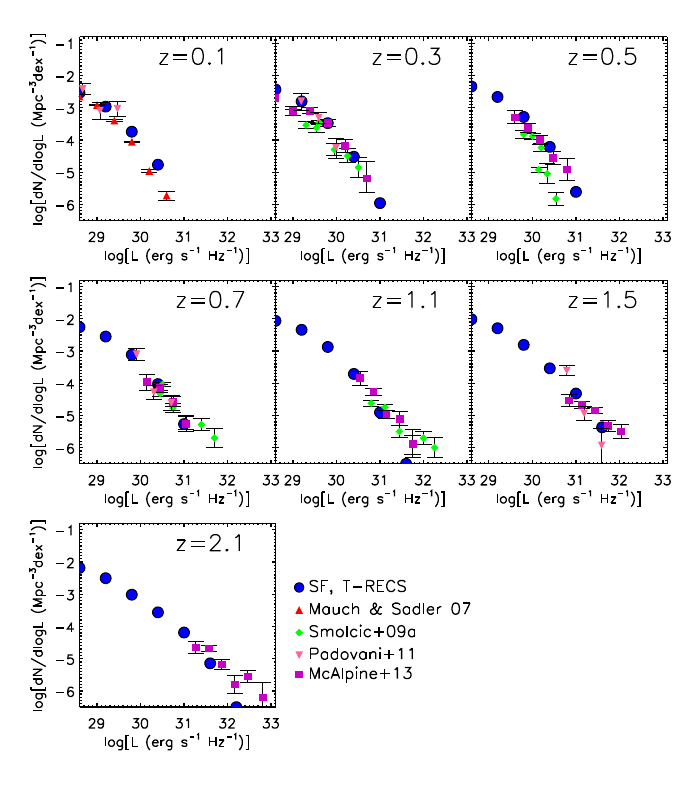}

  }
   \vspace{-0.5cm}
  \caption{Comparisons between our SFG 1.4\,GHz RLFs derived from the simulated catalogues
  (using the formalism described in Sect.~\ref{sec:SFG_model}) and observational determinations
  taken from literature (\citealt{Padovani2011}, \citealt{MauchSadler2007}, \citealt{Smolcic2009a} and \citealt{McAlpine2013}). The simulated sky area does not always allow sampling the highest luminosities for which we have observational determinations.}
  \label{fig:1d4_LF_SF}
\end{figure*}

\begin{figure*}
  \hspace{+0.0cm} \makebox[\textwidth][c]{
    \includegraphics[trim=4.5cm 2.0cm 5.0cm 0.0cm,clip=true,width=0.9\textwidth, angle=0]{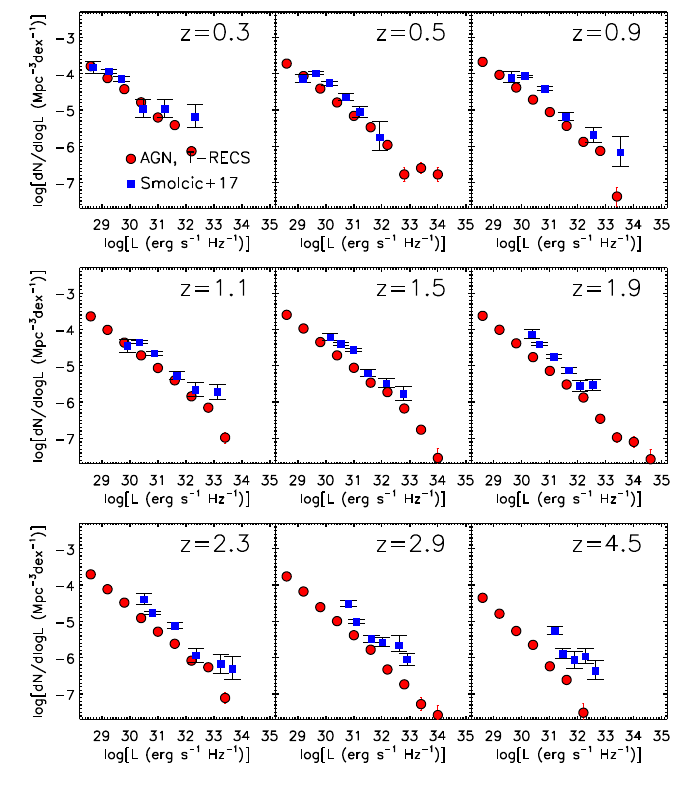}

  }
  \vspace{-0.5cm}
  \caption{Comparisons between our AGN 3\,GHz RLFs derived from the simulated catalogues (filled red circles)  using the formalism described in Sect.~\ref{sec:AGN_model}) and the \citet{Smolcic2017c} observational determinations (filled blue squares).}
  \label{fig:3_LF_AGN}
\end{figure*}

\begin{figure*}
\hspace{+0.0cm} \makebox[\textwidth][c]{
  \includegraphics[trim=4.5cm -0.2cm 5.0cm 0.0cm,clip=true,width=0.9\textwidth, angle=0]{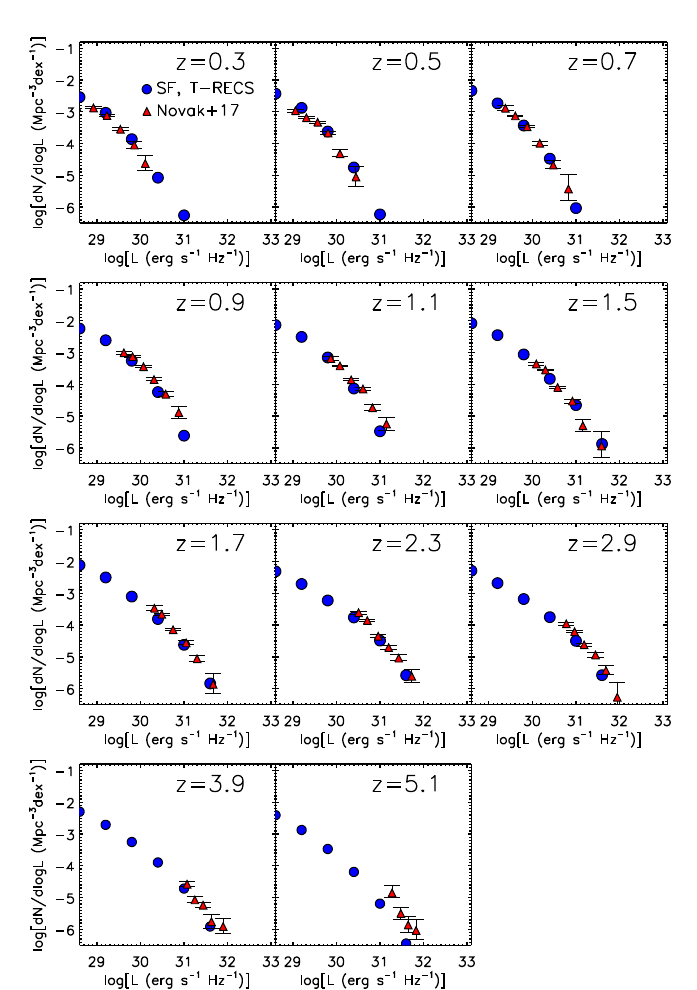}

  }
   \vspace{-0.5cm}
  \caption{Comparisons between our SFG 3\,GHz RLFs derived from the simulated catalogues
  (using the formalism described in Sect.~\ref{sec:SFG_model}) and the \citet{Novak2017} observational determinations.}
  \label{fig:3_LF_SF}
\end{figure*}

\begin{figure}
  \hspace{+0.0cm} \makebox[0.48\textwidth][c]{
    \includegraphics[trim=3.2cm 0.6cm 1.3cm 0.5cm,clip=true,width=0.48\textwidth, angle=0]{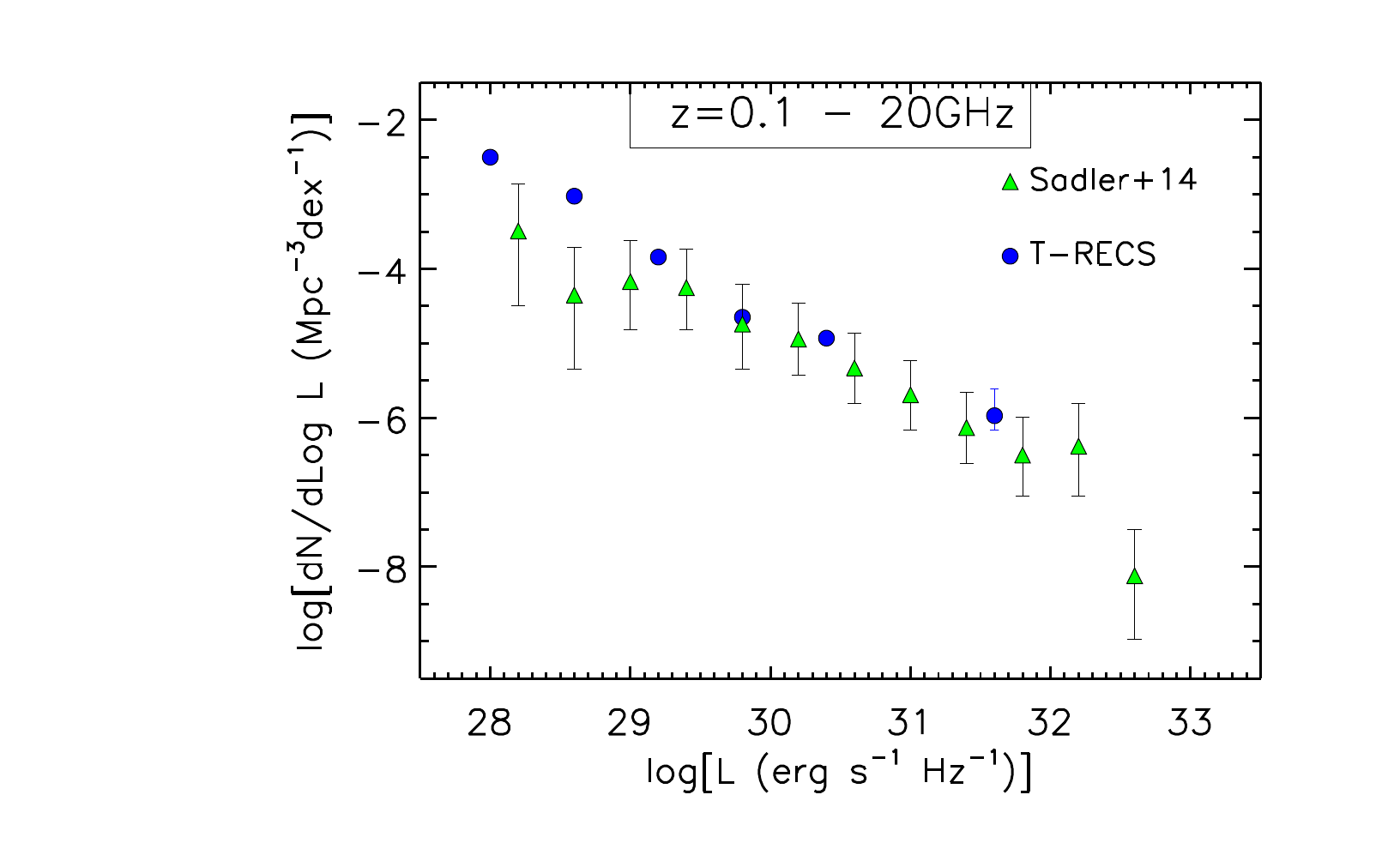}

  }
  \caption{Comparisons between our 20\,GHz RLFs of the whole (SFG+AGN) population
  derived from the simulated catalogues (using the formalism described in
  Sects.~\ref{sec:AGN_model} and ~\ref{sec:SFG_model}) and the \citet{Sadler2014} observational determinations. The simulated sky area does not allow sampling high luminosities.}
  \label{fig:20_LF_tot}
\end{figure}

\subsection{Polarized intensity}
We also include polarization information (in terms of the polarized intensity $P=\sqrt{Q^2+U^2}$) for each simulated source. For
SS-AGNs the polarized flux densities were generated by sampling
from the polarization fraction distribution at $1.4 \, \mathrm{GHz}$  by
\citet{Hales2014}. This distribution was found to be independent of flux
density down to a total intensity of $\sim$10\,mJy and perhaps even of 1\,mJy. In
the absence of better information, we have assumed that this distribution holds
at all frequencies.  

In the case of flat-spectrum sources we have exploited the high sensitivity
polarization measurements in seven bands (centred at 2.1, 5.5, 9, 18, 24, 33 and
38\,GHz) of a complete sample of 104 compact extragalactic radio sources
brighter than 200\,mJy at 20 GHz, carried out by \citet{galluzzi2018}. Again,
no indications of a flux-density dependence of the distribution of polarization
fractions was found. Hence polarized flux densities at 5.5, 9 and 18\,GHz
were assigned sampling the observed distributions and interpolating at
intermediate frequencies. The distribution at 1.4\,GHz was computed using the
polarization measurements by \citet{Condon1998} for a complete sample of 2810
flat-spectrum sources brighter than 200\,mJy at 20\,GHz, drawn from the
Australian Telescope Compact Array 20 GHz (AT20G) survey \citep{Murphy2010}.
Below 1.4\,GHz the polarization fraction of each source was kept constant at
the 1.4\,GHz value.

We used the polarization counts resulting from those frequency-dependent polarization fractions to compute effective spectral indices in polarization for our flat-spectrum sources, with the same method described in the previous section. Consistently with that analysis, we considered again the frequency intervals 1.4--4.8\,GHz and 4.8--20\,GHz and a dispersion of $\sigma=0.25$. 
This additional step allows our polarized sources to have a regular, more physical frequency spectrum, while still being consistent with different polarization fractions at different frequencies.

\subsection{Clustering}\label{sec:cluster_agn}
In order to assign AGN sources to underlying haloes of the base DM simulation
(see Sect. 1) and recover the correct spatial correlation functions we started
from \cite{janssen2012}, giving the fraction of galaxies hosting an RL
AGN as a function of the host galaxy stellar mass, $M_{\star}$.
\cite{janssen2012} model the LERG and HERG populations of AGN separately. For
the LERG population, the fraction is consistent with $f^{\rm LERG} \propto
M_{\star}^{2.5}$, and saturates at $M_{\star} > 10.^{11.6}\, M_{\odot}$; for
the HERG population, the dependence is shallower ($f^{\rm HERG} \propto
M_{\star}^{1.5}$).

We combined these results with the \cite{aversa2015} relation between stellar
mass and dark matter (DM) halo mass $M_\star = F(\mhalo)$ (their Table 2,
including redshift evolution) to compute  the probability that a halo of a
given mass $M_{\rm h}$ hosts a LERG or a HERG AGN. The probability distributions 
that we obtain peak at $\log (M_{\rm h}/M_{\odot})=12.5$ and 12.8 for LERG and HERG respectively, with a $\sigma$ widths of 0.4 and 0.3. 

We then mapped these two populations into our three observational categories:
\begin{itemize}
\item FSRQs  from the HERG population;
\item BL Lacs from the LERG population.
\item SS-AGNs morphologically classified as FR\,II/FR\,I (see Sect. \ref{sec:agn_size}) from the HERG/LERG population.
\end{itemize}

We acknowledge that, while there is broad overlap between the HERG/LERG and FR\,I/FR\,II classifications, there are also differences between the two classes. For example, a significant population of FR\,II LERGs exists \citep{Laing1994} and, while HERGs tend to have higher luminosities than LERGs, both HERGs and LERGs are found across the full luminosity range \citep[e.g.,][]{BestHeckman2012}. However, multifrequency evolutionary models for HERGs and LERGs as advanced as those adopted here don't exist yet. Also, clustering data on these source populations are still endowed with substantial uncertainties \citep[cf. ][and Sect.~\protect\ref {sec:valid_clust}]{Hale2018}; hence a sophisticated treatment of these populations is not warranted at this stage.

Each radio AGN in the simulated catalogue was associated to either the LERG or HERG mass distribution, as previously described, and a halo mass was drawn from it.  Finally,
the source was associated to the halo having the closest mass in the same
redshift bin. The source was given the exact redshift and coordinates of the
centre of the dark matter halo. Haloes already associated to a galaxy are excluded
from the list, thus preventing multiple associations.

\subsection{Source sizes}\label{sec:agn_size}

The different radio AGN populations have also  different morphologies.
According to the unified AGN model \citep[e.g.,][]{orr1982, antonucci1985,
netzer1985, netzer1987} the compact, typically unresolved, sources (FSRQs and
BL Lacs in our simulation) and the extended ones, typically exhibiting the
double-lobe morphology (SS-AGNs in our simulation) are described by the same parent
population, only viewed from a different angle from the jet axis.

\cite{dipompeo2013} re-examined the unified model by fitting the
distribution of intrinsic sizes of one parent population, using observational
size data \citep[e.g.,][]{barthel1989,singal2013} as a constraint. We used
their result as the base for our size modelling.

Depending on the population, we drew an intrinsic size from one of the distributions in \citet[][see their Table 2]{dipompeo2013}; sources
were then given a viewing angle $\theta$ with the usual uniform distribution in
$\sin(\theta)$, with different limiting angles. Specifically: 

\begin{itemize}
\item FSRQs and BL Lacs: intrinsic size from the ``empirically determined, narrow" distribution; viewing angle $0 < \theta \leq 5 \deg$;
\item steep-spectrum AGNs: intrinsic size from the ``Modelled Gaussians" distribution; viewing angle $5 < \theta \leq 90 \deg$.
\end{itemize}

This means that the apparent projected sizes of FSRQs and BL Lacs are
small, whilst steep-spectrum AGNs may be readily resolved by survey telescopes
with $\sim$ arcsecond resolution. Therefore, in our model, FR\,I/FR\,II
morphologies are associated to steep-spectrum sources.

%We discriminate between
We characterise FR\,I and FR\,II by means of the $R_{s}$ parameter, defined as the ratio
between the total projected source size and the projected distance between the
two bright hot spots, which typically occur either side of the core emission.  \cite{FanaroffRiley1974} traditionally classified 
sources with $R_{s} > 0.5$ as FR\,II, and with $R_{s} < 0.5$ as FR\,I; they also gave a boundary luminosity of $P=1.3 \times
10^{26}\,\hbox{W}\,\hbox{Hz}^{-1}$ at 178\,MHz between the two classes, which corresponds to $L_{\rm thr, 1.4\,GHz}=10^{25.4}\,\hbox{W}\hbox{Hz}^{-1}$ at 1.4\,GHz (for a spectral index of $-0.8$). 

\cite{lin2010} computed the distribution of $R_s$ for a sample of 1040
luminous, extended radio galaxies, and found it to be bimodal, thus reflecting
the FR\,I/FR\,II dichotomy. By using their $R_s$ distributions for the two
main classes of objects, and the luminosity threshold $L_{\rm thr, 1.4\,GHz}=10^{25.4}\,\hbox{W}\hbox{Hz}^{-1}$, we drew values of $R_s$ from a normal distribution with:
\begin{itemize}
\item $L_{\rm 1.4\,GHz}> 10^{25.4}\,\hbox{W}\hbox{Hz}^{-1}$: mean 0.62, rms 0.18
\item $L_{\rm 1.4\,GHz}<10^{25.4}\,\hbox{W}\hbox{Hz}^{-1}$: mean 0.17, rms 0.11.
\end{itemize}

Our scheme does not explicitly include the low-power, compact steep-spectrum population referred to as ``FR0" sources \citep[e.g.,][]{Sadler2014, Baldi2015, Baldi2016}. In fact, the available information on the luminosity function of these sources and on its evolution is insufficient to properly deal with them. However, the intrinsic size distribution by \citet{dipompeo2013} in principle embraces all source populations. In particular, the adopted size distribution for low-power sources includes a substantial fraction of objects with small observed sizes. From a more general point of view, the size distribution of sources with radio luminosity in the FR0 range is anyway dominated by star-forming galaxies, hence is little affected by a somewhat imprecise modelling of the AGN radio sources.

We note that the size we model is the total core+jet emission, therefore the largest extent 
that the AGN would have in the sky. The brightness distribution of each source, typically very complex, means that the actually measured size could be smaller. For steep-spectrum sources, the $R_s$ parameter can be used to scale the total size to that containing most of the brightness (the core and the hot spots). Flat-spectrum sources are typically core-dominated, and are typically not resolved even with sub-arcsecond VLBI observations.

%%%%%%%%%%%%%%%%%%%%%%%%%%%%%%
%FIGURES RELATED TO SECTION 5.2
\begin{figure*}

\includegraphics[width=8.5cm]{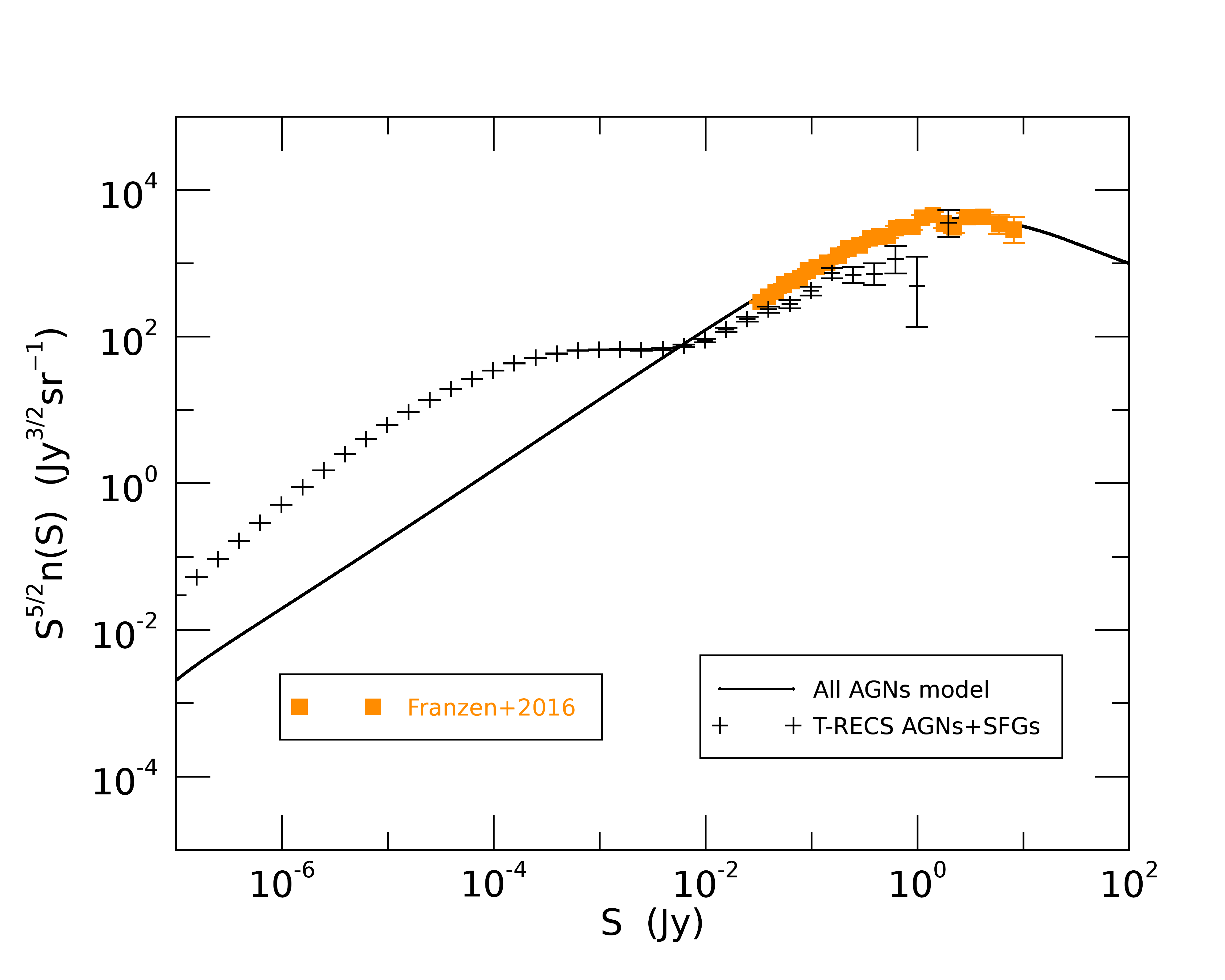}
\includegraphics[width=8.5cm]{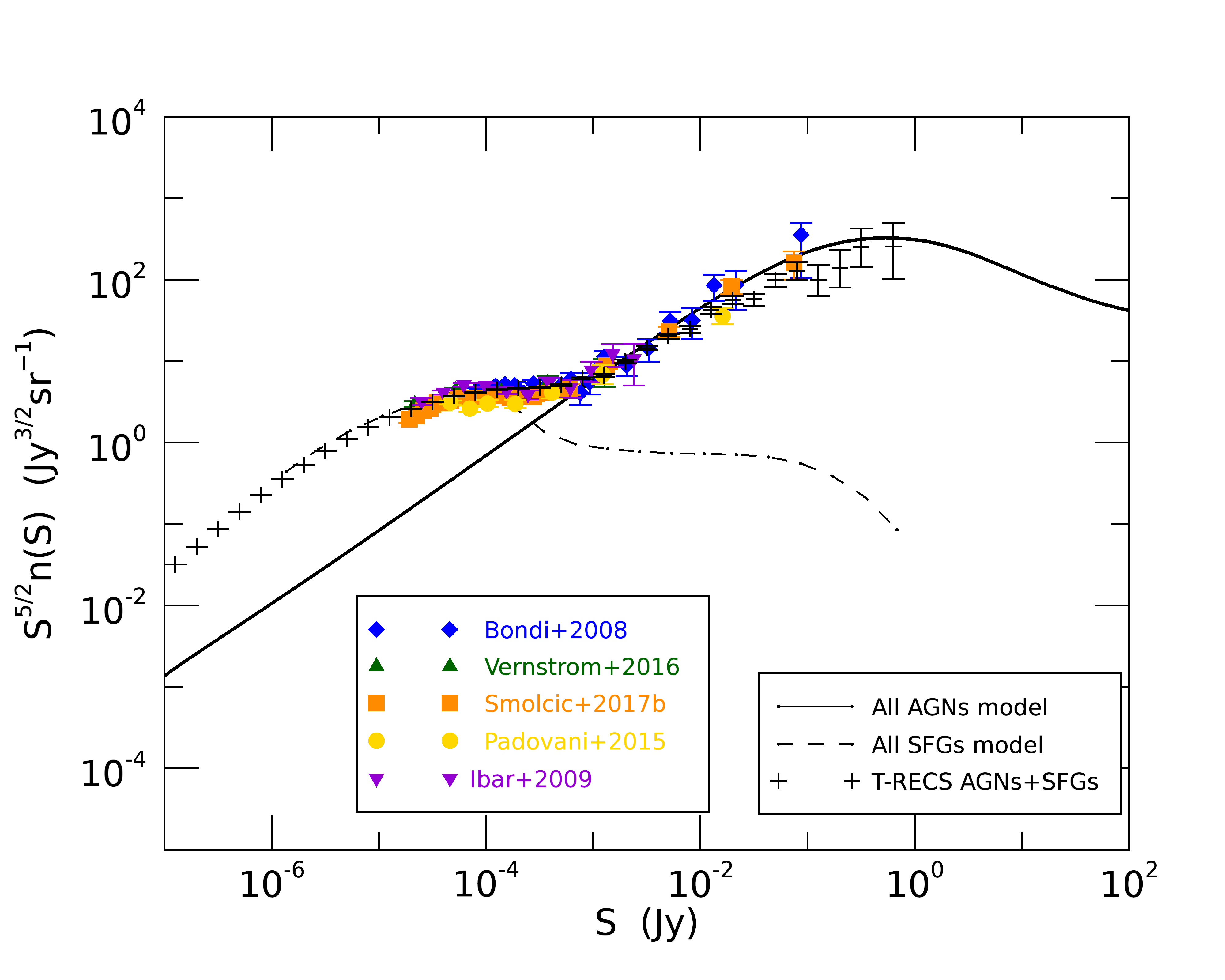}
\includegraphics[width=8.5cm]{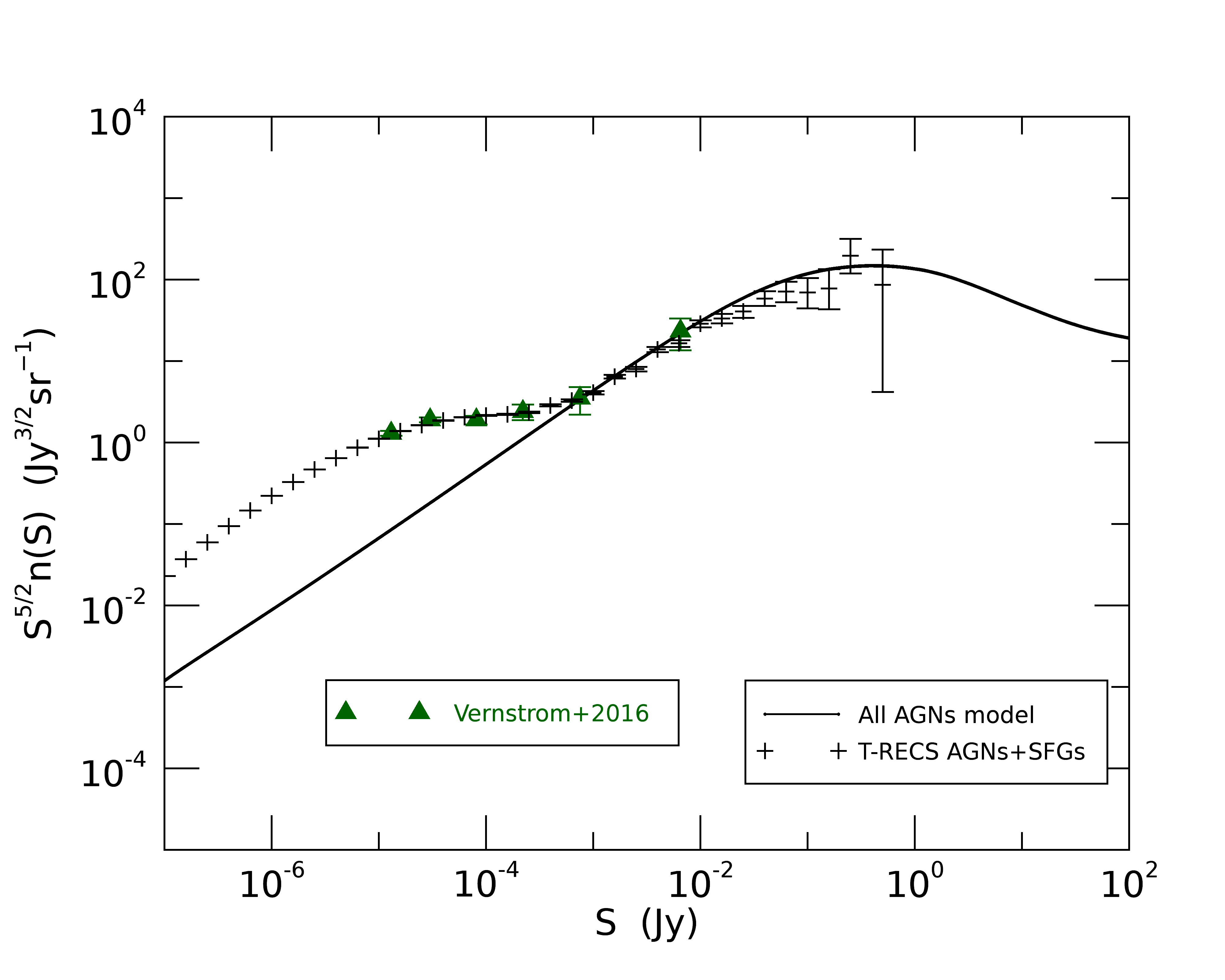}
\includegraphics[width=8.5cm]{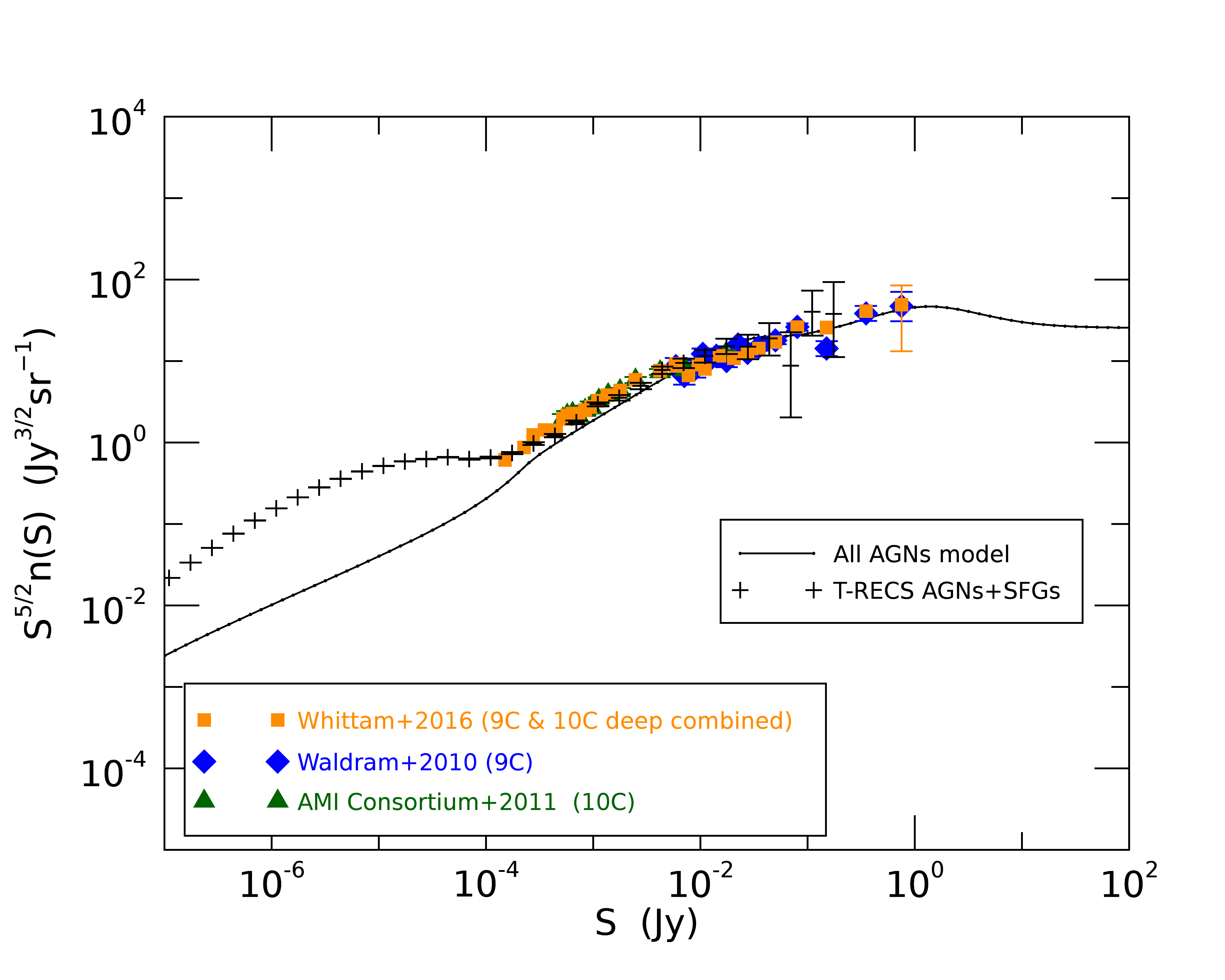}
\caption{Comparison of differential source counts in total intensity at 150\,MHz (top left) 1.4\,GHz (top right), 3\,GHz (bottom left) and 15\,GHz (bottom right) between T-RECS, models from \citet{Bonato2017} and the available data from \citet{franzen2016,Bondi2008,Vernstrom2016,Smolcic2017a,Padovani2015,ibar2009,whittam2016,waldram2010,davies2011}.}
\label{fig:tcounts} 
\end{figure*}

\begin{figure}
    \centering
    \includegraphics[width=8.5cm]{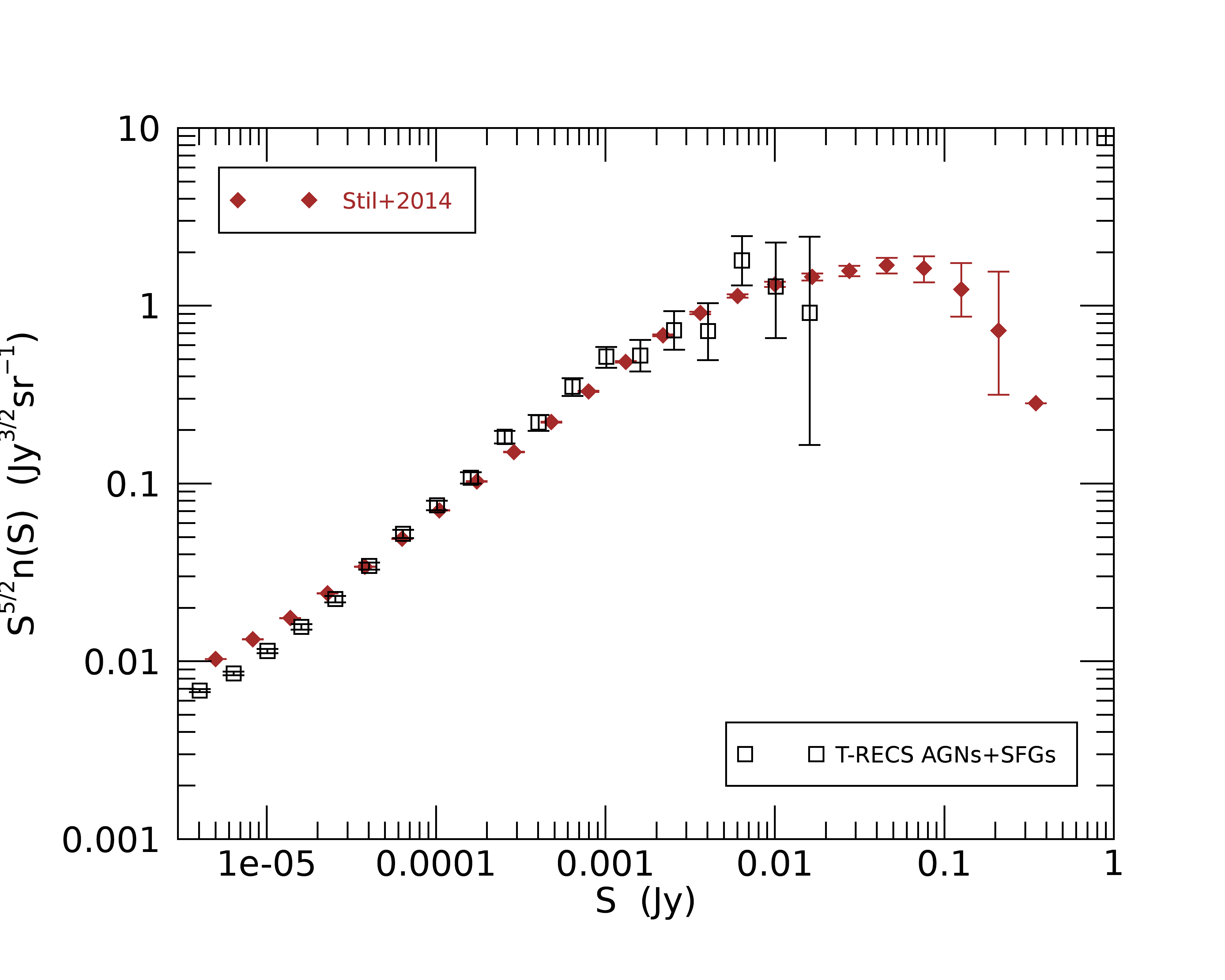}
    \caption{Comparison of polarization differential source counts from T-RECS with those estimated by \citet{stil2014}} 
    \label{fig:pcounts}
\end{figure}

%%%%%%%%%%%%%%%%%%%%%%%%%%%%%%
%%%%%%%%%%%%%%%%%%%%%%%%%%%%%%SFG

\section{Star-forming galaxies model description}\label{sec:SFG_model}
\subsection{Base evolutionary model}

The radio continuum emission of SFGs is tightly correlated with the star
formation rate \citep[SFR; e.g.][and references therein]{KennicuttEvans2012};
hence the redshift-dependent radio luminosity function of SFGs can be derived
from the evolving SFR function. A detailed study of the evolution of the SFR
function across the cosmic time was carried out by \citet{Cai2013} focussing
on IR data and by \citet{Cai2014} at $z\simgt 2$ focussing on UV and Ly$\alpha$
data but taking into account also dust attenuation and re-emission. The model
was extended by \citet{mancuso2015} and further successfully tested against
observational determinations of the H$\alpha$ luminosity function at several
redshifts. On the whole, data useful to derive the SFR function over
substantial SFR intervals are available up to $z\simeq 6$--7, with some
information extending up to $z\sim 10$ \citep[see also][]{aversa2015}. The
combination of dust extinction corrected UV/Ly$\alpha$/H$\alpha$ data with FIR
data yielded accurate determinations of the SFR function over such redshift
range.

The \citet{Cai2013} model also yields estimates of the effect of strong
gravitational lensing on the observed LFs of high-$z$ SFGs. We have exploited
it to take into account, in the simulations, strongly lensed (magnification
$\mu \ge 2$) galaxies. Although the contribution of these objects to the number
counts is small, they are a substantial fraction of the highest redshift
galaxies that should be detected by radio surveys at few $\mu$Jy/sub-$\mu$Jy
levels \citep{mancuso2015}.

The radio continuum emission of SFGs consists of a nearly flat-spectrum
free-free emission plus a steeper-spectrum synchrotron component. A calibration
of the relations between SFR and both emission components was derived by
\citet{Murphy2011} and \citet{Murphy2012}. Following \citet{mancuso2015} we
have rewritten such relations as follows:
\begin{eqnarray} \label{eq:Lff}
 L_{\rm ff}(\nu)&=&3.75\times 10^{19} \frac{\hbox{SFR}}{M_\odot/\hbox{yr}} \, \left(\frac{T}{10^4\,\hbox{K}}\right)^{0.3} \nonumber \\  &\cdot &\hbox{g}(\nu,\hbox{T})\,\exp{\left(-{h\nu\over k\hbox{T}}\right)}\ \hbox{W}\,\hbox{Hz}^{-1},
 \end{eqnarray}
where $T$ is the temperature of the emitting plasma (we have set $T=10^4\,K$)
and  $\hbox{g}(\nu,\hbox{T})$ is the Gaunt factor;
\begin{eqnarray}\label{eq:LsyncM}
\bar{L}_{\rm sync}&\simeq &1.9\times 10^{21} \left(\frac{\hbox{SFR}}{\hbox{M}_{\odot}\hbox{yr}^{-1}}\right) \left(\frac{\nu}{\hbox{GHz}}\right)^{-0.85} \nonumber \\
&\cdot& \left[1+\left({\nu\over 20\rm GHz}\right)^{0.5}\right]^{-1}\, \hbox{W}\,\hbox{Hz}^{-1}.
\end{eqnarray}
However, \citet{mancuso2015} showed that this relation, combined with
observational determinations of the local SFR function, leads to an
over-prediction of the faint end of the local radio LF of SFGs worked out by
\citet{MauchSadler2007}. A similar conclusion was previously reached by
\citet{massardi2010}.

Consistency with the \citet{MauchSadler2007} LF was recovered assuming that the
radio emission from low-luminosity galaxies is substantially suppressed,
compared to brighter galaxies. Following \citet{mancuso2015} we adopt a
deviation from a linear $L_{\rm sync}$--SFR relation described by:
\begin{equation}\label{eq:Lsync_nonlin}
L_{\rm sync}(\nu)=\frac{L_{\star,\rm sync}(\nu)}{\left({L_{\star,\rm sync}}/{\bar{L}_{\rm sync}}\right)^{\beta}+\left({L_{\star,\rm sync}}/{\bar{L}_{\rm sync}}\right)},
\end{equation}
where $L_{\star,\rm sync} = 0.886\, \bar{L}_{\rm sync}({\rm
SFR}=1\,M_\odot\,\hbox{yr}^{-1})$ with $\bar{L}_{\rm sync}$ given by
eq.~(\ref{eq:LsyncM}), and $\beta = 3$; at 1.4\,GHz, $L_{\star,\rm sync}\simeq
10^{28}\, \hbox{erg}\,\hbox{s}^{-1}$ $\hbox{Hz}^{-1}$. Again following
\citet{mancuso2015} we allow for a dispersion $\sigma_{\log L}=0.4$ around the
mean ${L}_{\rm sync}$--SFR relation.

\subsection{Total intensity number counts}
For our simulations, we sampled the redshift-dependent SFR functions by
\cite{mancuso2015} and converted each galaxy's SFR to a total radio emission in
the 150\,MHz--20\,GHz frequency range by taking into account synchrotron,
free-free, and thermal dust emission, as detailed in the following.

For free-free and synchrotron emission we use eqs.~(\ref{eq:Lff}) and
(\ref{eq:Lsync_nonlin}), respectively. Recent investigations
\citep{magnelli2015, Delhaize2017} have reported evidence of a weak but
statistically significant increase with redshift of the ratio between
synchrotron luminosity and far-infrared luminosity, generally believed to be a
reliable measure of the SFR, at least for $L_{\rm sync}\gg L_{\star,\rm sync}$
\citep[although the radio excess at high $z$ might be due to residual AGN
contributions;][]{Molnar2017}. We have allowed for the corresponding evolution
of the mean ${L}_{\rm sync}$--SFR relation adopting:
\begin{eqnarray}\label{eq:Magnelli}
\log L_{\rm synch,1.4\,\rm GHz}(z)&=&\log L_{\rm synch,1.4\,\rm GHz}(0) \nonumber \\
&+&2.35[1-(1+z)^{-0.12}].
\end{eqnarray}
\citet{Bonato2017} showed that this relation, based on the results of
\citet{magnelli2015}, yields a very good fit of the observational estimates of
the radio luminosity function of SFGs, recently determined up to $z\simeq 5$
\citep{Novak2017} as well as of the ultra-deep source counts at 1.4\,GHz
\citep{vernstrom2014, Vernstrom2016, Smolcic2017a}.

As illustrated by Fig.~5 of \citet{mancuso2015}, the rest-frame spectral energy
distribution (SED) of SFGs at $\nu \simgt 100\,$GHz is dominated by thermal
dust emission. This implies that, in the case of high-$z$ galaxies, this
component becomes important already at frequencies of a few tens of GHz in the
observer's frame. We have taken it into account using the \citet{Cai2013}
model.

The model comprises three SFG populations with different evolutionary
properties and different dust emission SEDs: `warm' and `cold' late-type
galaxies, and proto-spheroids. In the simulations, we have used the calibration
adopted by \citet{Cai2013} to translate its SFR into its total infrared (IR;
8--1000\,$\mu$m) luminosity, $L_{\rm IR}$:
\begin{equation}
\log(L_{\rm IR}/L_\odot)=\log(\hbox{SFR}/M_\odot\,\hbox{yr}^{-1})+9.892.
\end{equation}
The monochromatic luminosity of dust emission at the frequency $\nu$, $L_{\rm
dust}(\nu)$, was obtained from $L_{\rm IR}$ using the SED appropriate for each
population, given by \citet{Cai2013}, and added to the radio luminosity.

\subsection{Polarized intensity}

The polarization properties of star-forming galaxies are still poorly known.
The polarized signal is typically only a few percent of the total brightness,
but it depends strongly on frequency and on galaxy inclination, due to
depolarization effects. All these features are captured by \cite{sun_reich2012},
which study the polarization properties of Milky-Way-like galaxies with a
three-dimensional emission model. They derive polarization percentages as a
function of galaxy inclination for 5 frequencies: 1.4, 2.7, 4.8, 8.4 and
22\,GHz (see their Fig. 9). We model each curve with a fourth-order polynomial,
which we use to compute the polarization percentage at all 5 frequencies for a
randomly generated galaxy inclination $i$ (with a uniform distribution in
$\sin(i)$). This yields median polarization fractions of
$\sim$4.2\,\% at 4.8\,GHz and $\sim$0.8\,\% at 1.4\,GHz, which are consistent
with other observations \citep{test,taylor2014}. We then interpolate
linearly the polarization percentages for any other frequency. We finally
obtain a polarization spectrum by multiplying the frequency-dependent
polarization fraction by the total intensity spectrum. 
We used the interpolated polarization fraction values for each galaxy directly, 
without drawing them from a random distribution, to ensure that the resulting polarized spectrum is smooth in frequency. Differences in the spectra of individual sources are anyway obtained thanks to the dependence of polarization fractions on random inclinations and by the scatter in the total intensity spectral indices.

\subsection{Clustering} \label{sec:cluster_sfg}
In order to populate our simulated DM haloes (see Sect. 1) with radio-emitting
SFGs we use an abundance matching procedure \citep[e.g.][]{moster2013}.
Abundance matching is a method to constrain a relationship between two
quantities (in our case, radio luminosity $L$ and halo mass $M_{\rm h}$) whose individual 
distributions are known (radio luminosity
function and dark halo mass function).

We used the $L_{\rm SFR}$--$M_{\rm h}$ relation from \cite{aversa2015}, which
is of the form:
\begin{equation}
    L(M_{\rm h})= N \times \Big [\Big (\frac{M_{\rm h}}{M_b}\Big )^\alpha+\Big (\frac{M_{\rm h}}{M_b}\Big )^\omega \Big ]^{-1},\label{l(m)}
\end{equation}
where $N$, $\alpha$, $\omega$ and $M_b$ are free parameters which include
redshift evolution. We fitted for them separately for each redshift slice, by
requiring that the luminosity function derived from the mass function of the
cosmological simulation through eq. (\ref{l(m)}) matched as closely as possible
the radio luminosity function at 1.4\,GHz from \cite{Bonato2017}.

By inverting the best-fit $L(M_{\rm h})$ relation, we finally map radio
luminosities into halo masses for each redshift slice. This allows us to
associate galaxies to haloes in the light cone. Once a galaxy is associated to
a halo, we assign to it the redshift and the sky coordinates of the centre of
the halo.

As stated in Sect.~\ref{sec:dm_sim}, the minimum halo mass of the simulation is $M_{\rm h,min}=1.061 \times 10^8\,h^{-1}\,M_{\odot}$, which sets the minimum
luminosity of galaxies that we can associate to haloes with this method to
$L_{\rm min, 1.4 GHz}\simeq 10^{22}$\,erg/s/Hz, depending on the redshift. For less luminous galaxies, we assume a random distribution in the sky.

\subsection{Source sizes}

For SFG sizes we make reference to the scale radius $r_0$ of an exponential
emission intensity profile:
\begin{equation}
    I(r) = I_0 \exp(-r/r_0).
\end{equation}
\citet{shen2003} have given a relation between the optical half-light radius of
disk galaxies, $R$, and their stellar mass:
\begin{equation}
R({\rm kpc})=\gamma M_\star^\alpha (1+M_\star/M_0)^{\beta-\alpha},     \label{eqn:re}
\end{equation}
where $\alpha$, $\beta$, $\gamma$ and $M_0$ are free parameters. We have
performed a new fit for these parameters to match a set of radio observations:
those of \cite{biggs2006}, \cite{owen2008} and \cite{schinnerer2010}. These
three papers give galaxy sizes defined as the FWHM of a Gaussian intensity
profile: we derived the scale radius as $r_0 = 0.7\, r_{\rm FWHM}$. We
performed a joint fit to all three size distributions, finding the following
values of the parameters: $\alpha=0.115$, $\beta=0.898$, $\gamma=0.199$,
$M_0=3.016 \cdot 10^{10} M_{\rm sun}$.

For each source, we first computed the stellar mass $M_\star$ from the halo
mass $M_{\rm h}$ using the $M_{\star}$--$M_{\rm h}$ relation by
\cite{aversa2015}, with the parameter values listed in their Table 2. We then
computed the scale radius with eq.~(\ref{eqn:re}) with our values for
$\alpha$, $\beta$, $\gamma$ and $M_0$. We allowed for a dispersion of
$\sigma_{\rm ln R}=\sigma_2+(\sigma_1-\sigma_2)/[1+(M_*/M_0)^2]$, with
$\sigma_1=0.47$ and $\sigma_2=0.34$ as in \citet{shen2003}. Finally, the
physical size was converted into an apparent one, depending on the redshift.

Galaxy ellipticity for our galaxies was generated in terms of the $(e1,e2)$
components along the two main axes of the field of view. For the absolute value
of the ellipticity we used the distribution
\begin{equation}
    P(|e|)=|e|\left[\cos (\frac{\pi |e|}{2})\right]^2 \exp\left[-\frac{2|e|}{B}\right]^C
\end{equation}
with $B=0.19$ and $C=0.58$, as derived by \cite{tunbridge2016} performing shape
measurements on Very Large Array (VLA) Cosmological Evolution Survey
(COSMOS) radio data. We then generated random orientation
angles $0\le \theta \le 2\pi$ and projected the absolute ellipticity into the
two components with:
\begin{eqnarray}
e1&=&|e|\sin(\theta/2)\\
e2&=&|e|\cos(\theta/2).
\end{eqnarray}

%
%%%%%%%%%%%%%%%%%%%%%
\section{Validation}\label{sec:validation}
This section presents comparisons between the outputs of the T-RECS simulation
and the available real data.

\subsection{Luminosity functions}

Figures~\ref{fig:1d4_LF_AGN} and ~\ref{fig:1d4_LF_SF} compare the 1.4\,GHz
Radio Luminosity Functions (RLFs) of radio AGNs and SFGs, respectively, derived
from the simulated catalogues, with the observational determinations at several
redshifts available in the literature. The simulated catalogues were obtained
using the formalisms described in Sects.~\ref{sec:AGN_model} and
~\ref{sec:SFG_model}. For both populations, the figures show a very good
agreement between our simulated RLFs and literature data.

In Fig.~\ref{fig:3_LF_AGN} the 3\,GHz AGN RLFs derived from our simulated
catalogues are compared with the recent \citet{Smolcic2017c} observational
estimates derived from VLA-COSMOS. Note that, although the survey was carried out at 3\,GHz,
\citet{Smolcic2017c} presented RLFs converted to 1.4\,GHz. The conversion was
made using the measured spectral indices for sources ($\sim 50\%$ of the
sample) detected also at 1.4\,GHz. For the remaining $\sim 50\%$ of the sample
a constant spectral index $\alpha=-0.7$ was adopted.

We converted the \citet{Smolcic2017c} RLFs back to 3\,GHz using $\alpha=-0.7$.
Our simulated RLFs are consistent with these observational estimations up to $z
\sim 1.0$. At higher redshifts our RLFs are a factor of $\sim 3$-5 lower. This
discrepancy may be due to the fact that the \citet{Smolcic2017c} radio AGN
population includes galaxies hosting AGNs, irrespective of their radio
emission. ``Radio quiet'' AGNs are their dominant AGN sub-population at
$z\gtrsim 1.2$ (see their Fig. 7). According to the adopted model, the radio
emission of these objects is generally dominated by star formation in the host
galaxies. Therefore they are included in the SFG population. Moreover the
high$-$z \citet{Smolcic2017c} RLFs are significantly higher than the previous
observational determinations (see their Fig. 3).

Figure~\ref{fig:3_LF_SF} compares our 3\,GHz SFG RLFs with the recent
observational estimates by \citet{Novak2017}, again derived from the 3\,GHz
VLA-COSMOS survey. Also in this case we converted the \citet{Novak2017} RLFs,
tabulated at 1.4\,GHz in their paper, to 3\,GHz using a spectral index
$\alpha=-0.7$, as done by \citet{Novak2017} to make the opposite conversion
(for 75\% of their sources). The agreement with these observational results is
very good. Note that the space densities of SFGs are generally substantially
higher than those of radio AGNs, so that the contribution of radio quiet AGNs,
present in our RLFs but not in those by \citet{Novak2017}, does not make a
substantial difference.

Finally, in Fig.~\ref{fig:20_LF_tot}, we compare our simulated 20\,GHz RLFs of
the whole (AGN+SFG) population with the local \citet{Sadler2014} estimation.
Our results are consistent (within the error bars) with these data, apart at
the lowest radio luminosities shown in the figure, where the \citet{Sadler2014} RLFs may be
affected by incompleteness.

\subsection{Differential source counts}

Figure \ref{fig:tcounts} presents the comparison between T-RECS differential source counts in total intensity at 150\,MHz, 1.4\,GHz 3\,GHz and 15\,GHz  and with the available data \citep{franzen2016,Bondi2008,Vernstrom2016,Smolcic2017a,Padovani2015,whittam2016,waldram2010,davies2011,ibar2009}. We also checked the agreement at 20\,GHz with the AT20G \citep{Murphy2010} and at 610\,MHz with GMRT observations \citep{Garn2008}. Note that the simulated area
($5\times5\,\hbox{deg}^2$) is too small to adequately sample sources brighter
than a few hundred mJy at 1.4\,GHz. The shape of the counts yielded by our model is shown by the solid lines in Fig.~\ref{fig:tcounts}.
The total counts agree very well with the available data, both in the regime dominated by RL AGNs and in the one dominated by SFGs and RQ AGNs. We note the much closer agreement at 1.4\,GHz in the sub-mJy regime of our simulation with respect to \cite{wilman2008}, whose counts were around a factor 2 lower \citep{Bonaldi2016}. We also note that the good agreement persists over the whole frequency range explored, thus confirming the validity of the approaches used to associate frequency spectra to the sources, described in Sect. \ref{sec:AGN_model} and \ref{sec:SFG_model}.

Figure \ref{fig:pcounts} compares our polarization counts at 1.4\,GHz with the results from \cite{stil2014}. The counts were obtained from total intensity source counts after assuming a polarization fraction distribution obtained from NRAO VLA Sky Survey (NVSS) data. This analysis is therefore quite similar to what we have adopted for our simulation. Since our simulation in total intensity is consistent with the 1.4\,GHz data and the polarization fraction estimates from \cite{stil2014} are similar to those from \citet{Hales2014}, the agreement of our results with \cite{stil2014} is not surprising.

\subsection{Clustering}\label{sec:valid_clust}

Observational estimates of the 2-point angular correlation function,
$w(\theta)$, of both radio AGNs and SFGs have been recently obtained by
\citet{magliocchetti2017} and \citet{Hale2018}.

\citet{magliocchetti2017} investigated the clustering properties of a complete
sample of 968 radio sources brighter than 0.15 mJy at 1.4\,GHz, detected by the
VLA on the COSMOS field covering about $2\,\hbox{deg}^2$. Spectroscopic
redshifts are available for 52\% of the sources and photometric redshifts for a
further 40\%. Sources with redshift determinations were subdivided into radio AGNs (644
objects) and SFGs (247 objects) purely on the basis of their radio luminosity.
In practice, all sources with luminosity below/above a suitably chosen
redshift-dependent threshold were classified as SFGs/AGNs. 

\begin{figure}
\includegraphics[width=8.5cm]{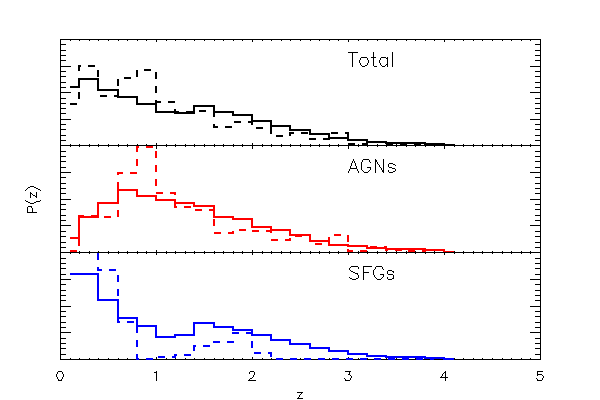}
\caption{Redshift distributions of \citet{magliocchetti2017} (dashed lines) compared with our catalogue for the same flux cut and sky area (solid lines).}
\label{fig:zdistr_maglio}
\end{figure}

The top panel Fig. \ref{fig:zdistr_maglio} shows that the global redshift distribution of simulated sources is consistent with the observational estimate, not surprisingly since the simulation reproduces reasonably well both the 1.4\,GHz counts (Fig.~\ref{fig:tcounts}) and the redshift-dependent luminosity functions of both the radio AGNs (Fig.~\ref{fig:1d4_LF_AGN}) and the SFGs (Fig.~\ref{fig:1d4_LF_SF}). Note that the true uncertainties of the observational determination are substantially larger than the Poisson fluctuations because of the contributions of the errors on photometric redshifts and of the sample variance.

As illustrated by the lower panels of Fig. \ref{fig:zdistr_maglio}, there are pronounced differences between the simulation and the estimates for each population.  %In particular, \citet{magliocchetti2017} have about a factor of two less SFGs than our simulation. 
Most of the difference is due to their selection of SFGs. For example, at $z=0.7$ they set the boundary between SFGs and radio AGNs at $\log(L_{1.4\rm GHz}/\hbox{erg}\,\hbox{s}^{-1}\,\hbox{Hz}^{-1})=30.5$, but the observed radio luminosity function of SFGs at this $z$ extends up to $\log(L_{1.4\,\rm GHz})\sim 31.8$ (see Fig.~\ref{fig:1d4_LF_SF}); at $z=1.5$ the boundary is at $\log(L_{1.4\,\rm GHz})\sim 31.3$ with the luminosity function reaching $\log(L_{1.4\,\rm GHz})\sim 32.2$, and so on. Thus the \citet{magliocchetti2017} criterion misses the brightest SFGs, especially around $z\sim 1$, where their redshift distribution has an unnatural minimum. The missed SFGs are classified as radio AGNs, resulting in the excess over the simulation around $z\sim 1$. This difference in the source classification between T-RECS and \cite{magliocchetti2017} needs to be taken into account when comparing the correlation functions. 

\begin{figure}
\includegraphics[width=8.5cm]{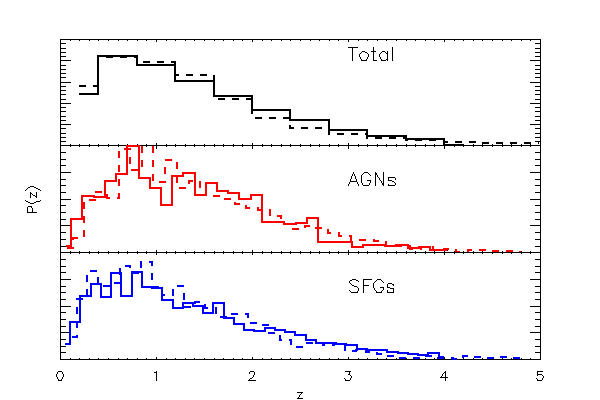}
\caption{Redshift distribution of the total sample of \citet{Hale2018} (dashed line) compared with our catalogue for the same flux cut and sky area (solid line).}
\label{fig:zdistr_hale}
\end{figure}

\begin{figure}
\includegraphics[width=8.5cm]{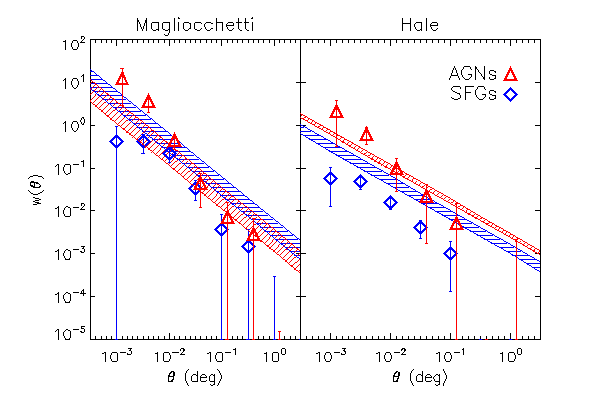}
\caption{Two-point angular correlation function $w(\theta)$ yielded by our simulation for radio AGNs and SFGs (points with error bars) compared to the results of \citealt{magliocchetti2017} (left) and \citealt{Hale2018} (right) for the same populations (shaded areas represent the best fit $\pm 1\,\sigma$ uncertainty in the normalization).}
\label{fig:wtheta}
\end{figure}

The broad minimum in the simulated redshift distribution of SFGs between $z=1$ and $z=1.2$ corresponds to the transition between the dominance of late-type galaxies, that are the main star forming population at  $z\le 1$, and proto-spheroidal galaxies that take over at higher $z$. This transition is in keeping with data on the age of stellar populations of the two galaxy types \citep[e.g.,][]{Bernardi2010}

\citet{Hale2018} measured $w(\theta)$ for the recently released, deeper 3\,GHz
VLA/COSMOS sample. They used a $5.5\,\sigma$ cut on the final catalogue,
corresponding to a mean flux density limit of $S_{3\,\rm GHz}\simeq 13
\,\mu\hbox{Jy}\,\hbox{beam}^{-1}$ or of $S_{1.4\,\rm GHz}\simeq 22
\,\mu\hbox{Jy}\,\hbox{beam}^{-1}$ for $S\propto \nu^{-0.7}$. Their catalogue
contains a total of 8928 sources over $\sim 2\,\hbox{deg}^2$.

The source identification was made by \citet{Smolcic2017b} by cross-matching
with optical, near-infrared, mid-infrared (Spitzer/IRAC) and X-ray data.

Optical counterparts to $\sim 86\%$ of the $\ge 5.5\,\sigma$ radio sources were found, and for 98\% of them photometric or spectroscopic redshifts were gathered. About 79\% of the sources were further classified as SFGs
or AGNs, based on various criteria, such as X-ray luminosity; observed
mid-infrared color; UV/far-infrared spectral energy distribution; rest-frame,
near-UV optical colour corrected for dust extinction; and radio excess relative
to that expected from the star formation rate of the hosts.

In interpreting the
\citet{Hale2018} results, it should also be taken into account that the
presence of an active nucleus detected via its IR/optical/UV/X-ray emission
does not necessarily imply that it gives a significant contribution to the
radio emission.

Figure \ref{fig:zdistr_hale} compares the redshift distributions from \cite{Hale2018} to those from our simulation, for the same flux density limit and sky area. The agreement is very good both on the total population and the AGN and SFG populations separately, thus indicating that source classification in this case is more consistent, which allows an easier comparison of the correlation functions as well.

Figure~\ref{fig:wtheta} compares the 2-point angular correlation
function, $w(\theta)$, for our AGNs and SFGs catalogues with those from
\cite{magliocchetti2017} and from \citet{Hale2018}. 
Both observational estimates adopted the standard power-law shape for the
correlation function: $w(\theta)=A\theta^{1-\gamma}$. Since the data did not
allow an accurate determination of both $A$ and $\gamma$ for each source
population, they fix $\gamma$ to 2 and 1.8, respectively, and fit for the normalization. As they point out, this implies that
the errors on $w(\theta)$ are underestimated.%\cite{magliocchetti2017} find $A_{\rm AGN}=1.6 \times 10^{-3}$ and $A_{\rm SFG}=4.3\times 10^{-3}$ and \cite{Hale2018} $A_{\rm AGN}=2.7\times 10^{-3}$ and $A_{\rm SFG}=1.3\times 10^{-3}$.

To produce the T-RECS
correlation functions, we used the full $5\times 5$\,deg$^2$ sky area and
the same flux limits of the observational estimates. We used the
\cite{1993ApJ...417...19H} estimator:
\begin{equation}
w(\theta)= \frac{DD \cdot RR}{DR \cdot DR} -1,
\end{equation}
where $DD$, $RR$ and $DR$ are the number of data-data, random-random and
data-random pairs separated by $\theta$. 

The random catalogue has been constructed redistributing uniformly the simulated sources between -2.5 and 2.5 degrees from the center of the patch, both in longitude and in latitude. The $w(\theta)$ was computed for 100 realizations. In Fig.~\ref{fig:wtheta} we show the mean values and their dispersions as a function of the angular scale.  

We find that, for T-RECS, implementation details on how  galaxies of a modelled halo mass $M_{\rm h}$ are associated to the actual haloes of the cosmological simulation have a non-negligible effect on the measured $w(\theta)$. The mass function of the cosmological simulation and that inferred from the luminosity function are somewhat different, which means that there is a deficit of suitable haloes in some mass ranges and a surplus in others. Allowing for some scatter between the predicted and the associated mass (as done for the results shown in Fig.~\ref{fig:wtheta}) alleviates the problem, however it typically favours association to smaller halo masses, given the shape of the mass function. 

Our simulations give amplitudes of the angular correlation function of SFGs consistently lower than those of both \citet{magliocchetti2017} and \citet{Hale2018}. The discrepancies are of 1.7\,$\sigma$ and 2.5\,$\sigma$ respectively, where $\sigma$ is the quadratic sum of errors of the observational estimates and of the simulations.  The higher amplitudes of the observationally estimated $w(\theta)$ imply higher average bias factors, i.e. higher halo masses. 
 
The halo masses inferred by \citet{magliocchetti2017} are far higher than those corresponding to their average stellar masses given by the halo to stellar mass relations by \citet{Moster2010}. According to these relations the \textit{minimum} halo mass of SFGs ($\log(M_{\rm min}/M_\odot)=13.1$) corresponds, at their average redshift ($\langle z\rangle=0.5$),  to a \textit{minimum} stellar mass of $\log(M_{\rm min}/M_\odot)=11.2$. But their \textit{average} stellar mass is $\log\langle M_{\star}/M_\odot\rangle =10.7\pm 0.5$, corresponding to an average halo mass of $\log(M_{\rm h}/M_\odot)\simeq 12.3$.
 
Looking at that the other way round, for the \citet{magliocchetti2017} flux cut the halo mass distribution of SFGs resulting from the T-RECS simulation peaks at $\log(M_{\rm h}/M_\odot)=12.5$, only slightly higher than the average halo mass corresponding to their average stellar mass. This small excess was indeed expected since, as argued above, \citet{magliocchetti2017} somewhat underestimated the fraction of high-$z$ SFGs, hence their mean stellar mass. 
 
The fact that the simulations yield halo mass distributions of SFGs consistent with  the stellar mass distributions derived by \citet{magliocchetti2017}, but falls short  of their estimates of the angular correlation function may suggest that the latter is anomalously high. \citet{Hale2018}  do not give estimates of stellar masses; hence the same test cannot be done. However, none of our simulations gives an amplitude of $w(\theta)$  as  large as that observationally estimated, suggesting that part of the discrepancy can be due to the issue, discussed above, of the association between galaxies and simulated haloes introducing some scatter in mass. 

The amplitudes of the AGN correlation functions from our simulations are consistent, on average, with the observational estimates by both \citet{magliocchetti2017} and \citet{Hale2018}, but the slope is somewhat steeper. We remind, however, that in both cases the data were not sufficient to simultaneously determine the amplitude and the slope  of the $w(\theta)$; hence $\gamma$ was not measured but was fixed at 2 and 1.8, respectively.
 
According to the simulations, the halo mass distribution of radio AGNs at the \citet{magliocchetti2017} flux density limit peaks at $\log(M_{\rm h}/M_\odot)=12.5$. At the mean redshift of 1.25 this corresponds to $\log(M_{\star}/M_\odot)=10.7$, in good agreement with the average stellar mass reported by \citet{magliocchetti2017}:  $\log\langle M_{\star}/M_\odot\rangle =10.9 \pm 0.5$.
 
Despite the differences noted above, and taking into account the uncertainties in both determinations, the agreement between the T-RECS clustering and the empirically-determined one is reasonably good.

\subsection{Source sizes}
\begin{figure}
\includegraphics[width=8.5cm]{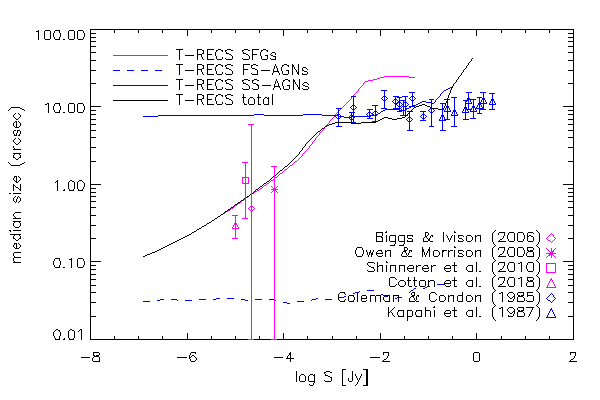}
\caption{Median sizes as a function of 1.4\,GHz flux for T-RECS (lines) and data from the literature \citep[][symbols with error bars]{schinnerer2010,biggs2006,owen2008,kapahi1987,coleman1985,cotton2018}.}
\label{fig:sizes}
\end{figure}

In Fig.\ref{fig:sizes} we compare the median source size as a function of 1.4\,GHz flux density from T-RECS with the literature. 

The size of AGNs does not depend on the flux density. As explained in Sec.\ref{sec:agn_size}, flat-spectrum (FS) sources (FSRQ and BLLac in our model) have been associated to small view angles and therefore to compact, beamed objects, while SS sources have been associated to larger view angles and FRI/FRII morphologies. This results in a marked difference (more than two orders of magnitude) between the size of flat and steep-spectrum AGNs. 
%There is a marked difference between the FS (FSRQ, BLLac) population and the SS one, which, following the unified AGN model, have been associated to objects having very small and large viewing angles, respectively. 
We find good agreement between T-RECS and observations \citep{coleman1985,kapahi1987} on the size of SS AGNs.

The size of SFGs increase with increasing flux, through the dependence of both quantities on the source redshift. To compare our results with the literature \citep{biggs2006,owen2008,schinnerer2010,cotton2018}, we show in Fig.\ref{fig:sizes} the median size at the median flux density of each observed sample, multiplied by 0.7 to convert from the Gaussian FWHM of the observational determinations to the exponential scale radius of our simulation.  Overall there is a good agreement between T-RECS sizes and observed sizes. 

The median size of the sample including all T-RECS sources (labelled ``total" in Fig. \ref{fig:sizes}) has a complex dependence on flux density, due to the fact that different populations dominate the total counts at different fluxes. At flux densities over $\sim$1\,mJy, the median size closely follows that of AGNs and it is therefore quite flat; at lower flux densities, the behaviour is that of SFGs and it is much steeper. 

\cite{}

%\newpage

\section{Available products}\label{sec:cats}
We release catalogues generated with the T-RECS code organised in three tiers:
\begin{enumerate}
\item \emph{deep}: 1\,deg$^2$ 1.4\,GHz flux limit 1\,nJy, data size 820\,Mb
\item \emph{medium}: 25\,deg$^2$, 1.4\,GHz flux limit 10\,nJy data size 6\,Gb
\item \emph{wide}: 400\,deg$^2$, 1.4\,GHz flux limit 100\,nJy, no clustering, data size 32\,Gb
\end{enumerate}
The format and the content of each catalogue is described in Appendix A. The frequencies at which we provide total intensity and polarization flux densities are listed in Table \ref{tab:cat_freqs}. They have been chosen to span the whole simulated 150\,MHz--20\,GHz frequency range, to be typically spaced by $\sim$ 30\,\% fractional bandwidth, and to include the frequencies allowing comparison with other data in Figures 1--7.

\begin{table}
\caption{Frequencies in the T-RECS released catalogues}
\begin{tabular}{l l l}
\hline
     Frequency &SKA bands & Data  \\
     
\hline
   150\,MHz &Low &(1)\\
   160\,MHz &Low& \\
   220\,MHz &Low& \\
   300\,MHz &Low&\\
   410\,MHz&Mid Band1&\\
   560\,MHz&Mid Band1& \\
   780\,MHz&Mid Band1& \\
   1.0\,GHz&Mid Band2& \\
   1.4\,GHz&Mid Band2&(2) (3) (4) (5) (7)\\
   1.9\,GHz&Mid Band3& \\
   2.7\,GHz&Mid Band3& \\
   3.0\,GHz&Mid Band4&(3)\\
   3.6\,GHz&Mid Band4& \\
   5.0\,GHz&Mid Band5a& \\    
   6.7\,GHz&Mid Band5a& \\
   9.2\,GHz&Mid Band5b& \\    
   12.5\,GHz&Mid Band5b& \\
   20.0\,GHz&&(6)\\
   \hline
\end{tabular}

\label{tab:cat_freqs}
\vspace{1mm}
 \scriptsize{
 $^1$Franzen et al. (2016) \\
 $^2$Bondi et al. (2008) \\
 $^3$Vernstrom et al. (2016) \\
 $^4$Smolcic et al. (2017) \\
 $^5$Padovani et al. (2015) \\
 $^6$Sadler et al. (2014)\\
 $^7$Ibar et al. (2009)} 
\end{table}

%and 20\,GHz. They have been chosen to being either the centre or minimum/maximum frequency of SKA Bands 1, 2, 5a and 5b; or the minimum/maximum frequency at which our simulation has been validated with data. 

More catalogues can be generated with different specifications if needed, by either requesting them or by running the T-RECS code to produce them.

Table~\ref{tab:intcounts} gives, as an example, the integral source counts at 3 frequencies (150\,MHz, 1.4\,GHz and 20\,GHz)
 computed from the 25\,deg$^2$ catalogue, which can be used to predict how many sources can be detected for a given flux density limit and source population. These numbers do not take into account the effect of PSF, noise or confusion, therefore they represent an ideal case. 

\begin{table*}
%\label{tab:intcounts}
\caption{Integral source counts $\log N(>S)$ at 150\,MHz, 1.4\,GHz and 20\,GHz computed from the 25\,deg$^2$ T-RECS catalogue, for all the modelled populations and sub-populations (identified by the numbers 1,2,3 for late-type, spheroidal and lensed spheroidal SFGs; 4,5,6 for FSRQ, BLLac and SS-AGNs). \label{tab:intcounts}}
\begin{tabular}{|c|c|cccc|cccc|}

 \hline
%\bf{1.4\,GHz}&Total&Tot AGNs&AGN Pop4&AGN Pop5&AGN Pop6&Tot SFGs&SFG Pop1 &SFG Pop2&SFG Pop3\\
\bf{150\,MHz}&&&&&&&&&\\
$\log S$\,[Jy]&$\log N(>S)$&&&&&&&&\\
&Total&Tot AGNs&AGN (4)&AGN (5)&AGN (6)&Tot SFGs&SFG (1) &SFG (2)&SFG (3)\\
\hline
%2.00& & & & & & & & & \\
%1.50& & & & & & & & & \\
%1.00& & & & & & & & & \\
0.50&2.72&2.72& & &2.72& & & & \\
0.00&3.39&3.39&2.11& &3.37& & & & \\
-0.50&3.90&3.88&2.81& &3.85&2.41&2.41& & \\
-1.00&4.39&4.37&3.11&2.11&4.34&3.15&3.15& & \\
-1.50&4.81&4.77&3.63&2.72&4.73&3.75&3.74&2.11& \\
-2.00&5.25&5.14&4.04&3.11&5.10&4.62&4.56&3.72&2.41\\
-2.50&5.82&5.47&4.45&3.63&5.42&5.57&5.43&5.01&3.39\\
-3.00&6.52&5.77&4.83&4.17&5.70&6.44&6.25&5.98&4.10\\
-3.50&7.19&6.06&5.21&4.58&5.98&7.15&6.97&6.69&4.71\\
-4.00&7.75&6.35&5.61&4.97&6.24&7.74&7.54&7.29&5.27\\
-4.50&8.22&6.65&5.99&5.36&6.51&8.21&7.98&7.81&5.76\\
-5.00&8.60&6.94&6.37&5.72&6.77&8.59&8.30&8.26&6.17\\
-5.50&8.90&7.24&6.75&6.08&7.02&8.89&8.53&8.63&6.53\\
-6.00&9.14&7.54&7.12&6.42&7.28&9.13&8.70&8.92&6.83\\
-6.50&9.34&7.84&7.47&6.76&7.53&9.32&8.83&9.15&7.08\\
-7.00&9.51&8.09&7.72&7.10&7.76&9.49&8.96&9.34&7.29\\
-7.50&9.65&8.22&7.83&7.35&7.88&9.63&9.10&9.47&7.42\\
-8.00&9.67&8.24&7.86&7.40&7.89&9.65&9.14&9.49&7.43\\

\hline
\bf{1.4\,GHz}&&&&&&&&&\\
$ \log S$\,[Jy]&$\log N(>S)$&&&&&&&&\\
&Total&Tot AGNs&AGN (4)&AGN (5)&AGN (6)&Tot SFGs&SFG (1) &SFG (2)&SFG (3)\\
\hline
% log S &logN at 1.4 GHz&&&&&&&&&\\
%\hline
%2.00& & & & & & & & & \\
%1.50& & & & & & & & & \\
%1.00& & & & & & & & & \\
%0.50& & & & & & & & & \\

%2.00& & & & & & & & & \\
%1.50& & & & & & & & & \\
%1.00& & & & & & & & & \\
%0.50& & & & & & & & & \\
0.00&2.11&2.11& & &2.11& & & & \\
-0.50&3.23&3.23&2.41& &3.15& & & & \\
-1.00&3.86&3.86&2.81& &3.82& & & & \\
-1.50&4.35&4.34&3.37&2.11&4.29&2.72&2.72& & \\
-2.00&4.80&4.78&3.82&2.72&4.73&3.34&3.34& & \\
-2.50&5.18&5.14&4.23&3.32&5.08&4.09&4.05&2.81&2.41\\
-3.00&5.61&5.49&4.61&3.91&5.41&4.98&4.88&4.31&2.72\\
-3.50&6.18&5.80&4.99&4.33&5.71&5.94&5.75&5.49&3.74\\
-4.00&6.84&6.10&5.37&4.74&5.99&6.75&6.54&6.32&4.40\\
-4.50&7.45&6.40&5.76&5.14&6.25&7.41&7.21&6.97&4.99\\
-5.00&7.98&6.70&6.13&5.50&6.52&7.95&7.74&7.54&5.51\\
-5.50&8.41&7.01&6.54&5.88&6.78&8.39&8.14&8.03&5.97\\
-6.00&8.76&7.30&6.89&6.20&7.03&8.75&8.43&8.46&6.37\\
-6.50&9.06&7.63&7.29&6.58&7.29&9.04&8.66&8.80&6.71\\
-7.00&9.31&7.92&7.60&6.89&7.54&9.29&8.85&9.09&7.01\\
-7.50&9.55&8.15&7.81&7.25&7.78&9.53&9.04&9.35&7.29\\
-8.00&9.67&8.24&7.86&7.40&7.89&9.65&9.14&9.49&7.43\\

\hline
\bf{20\,GHz}&&&&&&&&&\\
$\log S$[Jy]&$\log N(>S)$&&&&&&&&\\
&Total&Tot AGNs&AGN (4)&AGN (5)&AGN (6)&Tot SFGs&SFG (1) &SFG (2)&SFG (3)\\
\hline

-1.00&2.81&2.81&2.41& &2.59& & & & \\
-1.50&3.64&3.64&2.89& &3.56& & & & \\
-2.00&4.24&4.23&3.46&2.41&4.14&2.41&2.41& & \\
-2.50&4.70&4.69&3.93&3.07&4.60&3.07&3.07& & \\
-3.00&5.13&5.12&4.37&3.58&5.02&3.69&3.66&2.41&2.11\\
-3.50&5.52&5.44&4.75&4.04&5.32&4.71&4.56&4.15&2.81\\
-4.00&6.03&5.72&5.17&4.52&5.54&5.73&5.46&5.39&3.73\\
-4.50&6.70&6.03&5.58&4.96&5.78&6.60&6.32&6.27&4.41\\
-5.00&7.35&6.38&6.01&5.38&6.05&7.30&7.05&6.94&4.99\\
-5.50&7.91&6.73&6.43&5.78&6.32&7.88&7.64&7.51&5.53\\
-6.00&8.39&7.10&6.86&6.18&6.59&8.36&8.09&8.02&6.00\\
-6.50&8.78&7.45&7.24&6.54&6.85&8.76&8.44&8.47&6.41\\
-7.00&9.12&7.76&7.57&6.88&7.10&9.10&8.72&8.86&6.78\\
-7.50&9.41&7.99&7.78&7.21&7.35&9.40&8.96&9.19&7.13\\
-8.00&9.66&8.13&7.85&7.39&7.60&9.65&9.14&9.48&7.43\\

\hline
\end{tabular}
\end{table*}
%[This could also be an appendix]
%\subsection{Simulated catalogues}
%\subsection{Simulated skies}
%\subsection{User-generated catalogues}

%\section{Future developments}
%\subsection{wide area }
%\subsection{radio-quiet AGNs}
%\subsection{optical counterparts?}
%\subsection{lensing?}

\section{Conclusions}\label{sec:conclu}
We have presented a new simulation of the continuum radio sky, the Tiered Radio Extragalactic Continuum Simulation (T-RECS).  The main goal of this simulation is to allow the production of mock deep radio data. In the context of the SKA, those mock observations could be used to test the validity of scientific proposals, optimise survey design or test data analysis methods in advance of the real data. Our outputs\footnote{the catalogs are available at CDS via anonymous ftp to cdsarc.u-strasbg.fr (130.79.128.5) or via http://cdsarc.u-strasbg.fr/viz-bin/qcat?VII/282} and code \footnote{the code is available on github: https://github.com/abonaldi/TRECS.git} are released publicly.
Our simulation models two main radio-source populations: AGNs, further divided into FSRQ, BL Lac and SS-AGNs; and SFGs, further divided into warm$+$cold late-type galaxies, spheroids and lensed spheroids. 

Our model for the source continuum spectra
holds for a very wide frequency range, from 150 MHz to $\sim 20$\,GHz. For the AGN population, this has been achieved by allowing the sources to have a different spectral index below and above $\sim$5\,GHz, constrained by the modelled counts from \cite{massardi2010} and \cite{DeZotti2005} respectively for the lower and higher  frequency range. For the SFG population, our spectral modelling includes synchrotron, free-free and thermal dust emission, all expressed as a function of SFR \citep{mancuso2015,Bonato2017,Cai2013}. Our outputs in total intensity are shown to be in very good agreement with all most recent data compilations (luminosity functions and differential source counts) at several frequencies between 150\,MHz and 20\,GHz. 

Our polarization models are based on polarization fractions derived from observations  \citep{Hales2014,galluzzi2018} and from emission models \citep{sun_reich2012} for the AGN and SFG populations, respectively; they reproduce extremely well the polarization differential source counts estimated by \cite{stil2014}. We provide polarized intensity for the sources ($P=\sqrt{Q^2+U^2}$), which can be turned into $Q$ and $U$ once a polarization angle is assumed. 

We simulated clustering by modelling the mass properties of our populations and associating galaxies to dark matter haloes of a high-resolution cosmological simulation \citep[P-millennium,][]{baugh2017}. Our 2-point correlation functions have been successfully compared to the recent observational determinations from \cite{magliocchetti2017} and \cite{Hale2018}. The size of the cosmological simulation box (boxes of side 800 Mpc/h) constrains the size of the FoV for the simulation including clustering to $5 \times 5$\,deg$^2$.

%Finally, our catalogue include shape and size information. 
Our catalogue includes shape and size information, which can be used to generate images of the FoV, for example with the \emph{galsim} package. 

This is the first release of our simulation; in further releases, we plan to include an explicit modelling of the RQ AGN population;  modelling clustering on larger angular scales, therefore allowing to simulate larger area surveys; include the effect of weak gravitational lensing by distorting the ellipticity of galaxies according to a shear field. We also plan to update and improve our models to keep them in good agreement with new data when they become available.

\appendix
\section{Content of the catalogues}
The T-RECS outputs are separate catalogues for the AGN and SFG populations,
with the number and content of columns varying slightly between the two, as a
result of the different modelling.
%{\bf CHECK THIS AFTER VIZIER UPLOAD Both catalogues are fits binary tables, which can be read by standard packages. The data is embedded in a structure, and tag names are associated to each column.} 
To ease
the interpretation of results, together with the observable quantities that a
real data catalogue would typically contain (coordinates, redshift, flux
density, shape, etc.) we also included some other key quantities that are not
readily observable but are important components of the model (e.g., the dark
mass associated to each galaxy, the intrinsic luminosity and the SFR).
The columns of the AGN and SFG catalogues are listed in Tables A1 and A2, respectively.

The catalogues are available in two formats: fits binary table and ASCII. The ``deep" and the ``medium" catalogues include two files each, one for SFGs and one for AGNs. The ``wide" catalogue consists of 14 files: one for the AGN population and 13 for SFGs. This has been done to reduce the size of the single files to $\sim$2\,Gb each. 

\begin{table*}
\label{tab:agncat}
\begin{tabular}{m{1.5cm} m{2.5cm} m{2.5cm} m{9cm}}
\hline
Column &Tag Name &Units &Description\\
\hline
1&Lum$_{1400}$&log(erg/s/Hz)&Luminosity at 1.4 GHz\\
2-19&I$_{\rm freq}$&mJy&Total intensity flux density of the source at frequency \emph{freq} for the frequencies listed in Table \ref{tab:cat_freqs} \\
20-37&P$_{\rm freq}$&mJy&Polarized flux density of the source at frequency \emph{freq} for the frequencies listed in Table \ref{tab:cat_freqs}\\
38&Mh&log($M_{\rm sun}$)&Dark halo mass\\
39&x\_coord&degs&First angular coordinate for the flat-sky approximation  (see end of Sect. 1 for more details)\\
40&y\_coord&degs&Second angular coordinate for the flat-sky approximation (see end of Sect. 1 for more details)\\
41&latitude&degs&Latitude spherical coordinate for a chosen centre of the field\\
42&longitude&degs&Longitude spherical coordinate for a chosen centre of the field\\
43&redshift&&redshift\\
44&phys size&Kpc&Physical length of the core+jet emission\\
45&angle&degrees&Viewing angle between the jet and the line-of-sight\\
46&size&arcsec&Projected apparent size of the core+jet emission\\
47&Rs&&Ratio between the distance between the spots and the total size of the jets, for the FR I /FR II classification. Non null only for steep-spectrum sources (see section for more details)\\
48&PopFlag&&Number identifying the sub-population: 4, 5, 6 for FSRQ, BL Lac and SS-AGNs, respectively.\\
\hline
\end{tabular}
\caption{Structure of the AGN catalogues released with this paper. Catalogues produced with the T-RECS code will have the same format except for the number and list of frequencies and the possibility to optionally output the luminosities for each frequency as additional columns.}
\end{table*}

\begin{table*}
\label{tab:sfgcat}
\begin{tabular}{m{1.5cm} m{2.5cm} m{2.5cm} m{9cm}}
\hline
Column &Tag Name &Units &Description\\
\hline
1&logSFR&log($M_{\rm sun}$)/yr&SFR\\
2:19&I$_{\rm freq}$&mJy&Total intensity flux density of the source at frequency \emph{freq} for the frequencies listed in Table \ref{tab:cat_freqs}\\
20:37&P$_{\rm freq}$&mJy&Polarized flux density of the source at frequency \emph{freq} for the frequencies listed in Table \ref{tab:cat_freqs}\\
38&Mh&log($M_{\rm sun}$)&Dark halo mass\\
39&x\_coord&degs&First angular coordinate for the flat-sky approximation  (see end of Sect. 1 for more details)\\
40&y\_coord&degs&Second angular coordinate for the flat-sky approximation (see end of Sect. 1 for more details)\\
41&latitude&degs&Latitude spherical coordinate for a chosen centre of the field\\
42&longitude&degs&Longitude spherical coordinate for a chosen centre of the field\\
43&redshift&&redshift\\
44&size&arcsec&Projected apparent size of the disc\\
45&e1&&First ellipticity component\\
46&e2&&Second ellipticity component\\
47&PopFlag&&Number identifying the sub-population: 1, 2, 3 for late-type, spheroidal and lensed spheroidal galaxies, respectively.\\
\hline
\end{tabular}
\caption{Structure of the SFG catalogues released with this paper. Catalogues produced with the T-RECS code will have the same format except for the number and list of frequencies and the possibility to optionally output the luminosities for each frequency as additional columns.} 
\end{table*}

\section*{Acknowledgments}

We thank the referee for useful suggestions which improved the quality of the paper. 

GDZ acknowledges financial support from ASI/INAF agreement n.~2014-024-R.1 for the {\it Planck} LFI Activity of Phase E2, from the ASI/Physics Department of the university of Roma€"Tor Vergata agreement n. 2016-24-H.0 for study activities of the Italian cosmology community.

AB and SK acknowledge the use of the Cosmology Machine (COSMA) in Durham University, and the help of J. Helly and the COSMA support team, for accessing and processing the outputs of the P-Millennium cosmological simulation.   

\bibliography{trecs}
\bibliographystyle{mn2e_plus_arxiv}

\bsp

\label{lastpage}

\end{document}